\documentclass[sn-mathphys-num]{sn-jnl}

\usepackage{times}  
\usepackage{helvet} 
\usepackage{multirow}
\usepackage{courier}  
\usepackage{tikz}
\usetikzlibrary{shapes.geometric, arrows, fit, positioning} \usetikzlibrary{decorations.pathreplacing}
\usepackage{pgfplots}

\usepgfplotslibrary{fillbetween}

\usepackage{tcolorbox}
\usepackage{graphicx}  
\usepackage{pifont}

\newcommand{\full}[1]{#1}
\newcommand{\short}[1]{}
\newcommand{\forarxiv}[1]{#1}
\def\shortcite{\citeyearpar}

\usepackage{amsmath,amsthm,amsfonts}
\usepackage{microtype}
\usepackage{nicefrac}
\usepackage{enumitem}  

\DeclareMathOperator*{\argmin}{argmin}

\newcount\Comments  
\definecolor{darkgreen}{rgb}{0,0.6,0}

\def\tREMJ{\tau\shyp\REMJ}
\def\tREIMJ{\tau\shyp\REIMJ}
\def\tREMD{\tau\shyp\REMD}
\def\tREPL{\tau\shyp\REPL}

\newcommand{\insupp}[1]{}
\newcommand{\mypara}[1]{\vspace{-0.3mm}\paragraph{#1}}
\newcommand{\labeq}[2]{\begin{equation}\label{eq:#1}#2\end{equation}}
\newcommand{\kibitz}[2]{\ifnum\Comments=1{\color{#1}{#2}}\fi}

\newcommand{\rmr}[1]{\kibitz{blue}{[Reshef says:#1]}}

\def\calR{\mathcal{R}}
\def\calG{\mathcal{G}}
\def\calB{\mathcal{B}}
\mathchardef\shyp="2D
\def\SM{\textit{SMJ}}
\def\MJ{\textit{MJ}}
\def\PL{\textit{PL}}
\def\IMJ{\textit{IMJ}}

\def\MN{\textit{MN}}
\def\MD{\textit{MD}}

\def\RE{\textit{SQ}}
\def\REMJ{\textit{SQ-MJ}}
\def\REIMJ{\textit{SQ-IMJ}}
\def\REMD{\textit{SQ-MD}}
\def\REMN{\textit{SQ-MN}}

\def\REPL{\textit{SQ-PL}}
\def\tSM{\tau\shyp\SM}
\def\tSQMJ{\tau\shyp\REMJ}
\def\tREIMJ{\tau\shyp\REIMJ}
\def\tSQMD{\tau\shyp\REMD}
\def\tREMN{\tau\shyp\REMN}

\def\tREPL{\tau\shyp\REPL}
\def\tRER{\tau\shyp\RE\shyp\calR}

\def\eps{\varepsilon}
\def\ol{\overline}

\def\range{\gamma}
\def\safe{\alpha}
\def\live{\beta}
\newcommand{\floor}[1]{\left\lfloor#1\right\rfloor}

\newcommand{\sect}[1]{\vspace{-2mm}\section{#1}\vspace{-1mm}}
\newcommand{\subsect}[1]{\vspace{-1mm}\subsection{#1}\vspace{-0.7mm}}

\newtheorem{theorem}{Theorem}
\newtheorem{corollary}[theorem]{Corollary}
\newtheorem{proposition}[theorem]{Proposition}

\newtheorem{observation}[theorem]{Observation}
\newtheorem{lemma}[theorem]{Lemma}

\newtheorem{definition}{Definition}
\newtheorem{example}{Example}
\newtheorem{remark}{Remark}

\usepackage{xcolor}
\newcommand{\revision}[1]{#1}

\begin{document}
\title[Safe Voting: Resilience to Abstention and Sybils]{Safe Voting: Resilience to Abstention and Sybils\footnote{This full paper combines and extends two preliminary papers published in conferences; in particular, the work of Shahaf et al.~\cite{srsc} and its follow-up by Meir et al.~\cite{srscpp}. This version provides more extensive discussions, a presentation that combines these two preliminary papers, and additional results.}}
\author[1]{\fnm{Reshef} \sur{Meir}}\email{reshefm@technion.ac.il}
\affil*[1]{\orgname{Technion---Israel Institute of Technology}}
\author[2]{\fnm{Gal} \sur{Shahaf}}\email{gal.shahaf@mail.huji.ac.il}
\affil*[2]{\orgname{Independent researcher}}
\author[3]{\fnm{Ehud} \sur{Shapiro}}\email{udi.shapiro@gmail.com}
\affil[3]{\orgname{Weizmann Institute of Science}}
\author*[4]{\fnm{Nimrod} \sur{Talmon}}\email{talmonn@bgu.ac.il}
\affil[4]{\orgname{Ben-Gurion University}}

\abstract{Voting rules may implement the will of the society when all eligible voters vote, and only them. However, they may fail to do so when sybil (fake or duplicate) votes are present and when only some honest (non sybil) voters actively participate. As, unfortunately, sometimes this is the case, our aim here is to address social choice in the presence of sybils and voter abstention.
To do so, we build upon the framework of Reality-aware Social Choice: we assume the status quo as an ever-present distinguished alternative, and study \emph{status quo Enforcing (QUE) voting rules}, which add virtual votes in support of the status quo. We characterize the tradeoff between \emph{safety} and \emph{liveness} (the ability of active honest voters to maintain/change the status quo, respectively) in several domains, and show that the  voting rules are often optimal. \revision{Our characterization identifies the exact conditions under which mechanisms remain both resilient to sybils and responsive to verified participation, offering a quantitative tool for designers to measure the benefit of increased participation and verified identities. 
}
We comment on the applicability of our methods and analyses to the governance of digital communities.}
\maketitle





\sect{Introduction}\label{section:introduction}

Voting procedures are a simple and widely used way to aggregate the preferences of multiple individuals. Voting, however, can truly reflect the will of the society only insofar as all eligible people in the society---and only them---vote. 


Indeed, this corresponds to two different challenges: the problem of \textbf{sybil votes} and the problem of \textbf{partial participation}. These problems are particularly crucial as a single vote may tilt a majoritarian group decision and as such, sybils infiltrating a group of agents that employ egalitarian democratic group decision making literally pose an existential threat to the group; a threat that 
is further amplified in the presence of vote abstention.


While the extensive research on sybil identification (see in Section~\ref{section:related work}) may help keep the fraction of sybils in such communities in check, in online communities one can never assume sybils to be perfectly identified and completely eradicated. 
Thus, our goal in this context is to enhance social choice theory with effective group decision mechanisms for communities with bounded sybil penetration; put differently, to develop group decision making processes that can be safely used in online communities that are not sybil-free.

Orthogonally, the problem of partial participation in online voting is particularly acute, as online voting often exhibit very low participation rates~\cite{christian2005voting,jonsson2011user}. For example, in the 2006 Cambridge MA participatory budgeting program, only 7.5\% out of $\sim$64000 eligible voters actually participated~\cite{cambridgePB} (see Section~\ref{section:related work}). A recent position paper has argued that low participation effectively invalidates nearly all theoretical results in social choice, and called the community to study and mitigate the effects of abstention~\cite{meir2025tyranny}.  
The current paper has a more ambitious goal (albeit in the restricted context of simple decision rules), namely to handle both sybil votes and vote abstention with the single tool.



\paragraph{Voting with a status quo}
Our key approach takes the current state of affairs as a special reference alternative, following the \emph{Reality-Aware Social Choice} framework~\cite{realsoc}: we use the status quo as an anchor to guarantee both \emph{safety} -- the inability of sybils or abstentions to change the status quo against the will of the genuine agents; and \emph{liveness} -- the ability of the genuine agents to change the status quo.

\revision{
Our parametrized \emph{Status Quo Enforcing (SQE) mechanism} is is simple and general: it takes some base voting rule with desired properties, but adds a fraction $\tau$ of additional `virtual voters' who vote for the status quo.  We argue that this straight-forward modification provides a good (often optimal) tradeoff between safety and liveness.

\paragraph{Status Quo as instrumental conservatism}
%
We emphasize that the bias toward the status quo in our framework does not reflect any normative endorsement of the current state of affairs, or an identification of radical changes with  catastrophe (in contrast to Burke and other conservative thinkers, see \cite{conservatism}). Rather, it stems from the epistemic fact that the status quo is the only alternative whose legitimacy is certain and publicly verifiable, whereas changes are prone to \emph{costly errors}. This is similar to the instrumentalist argument in favor of conservatism made by Vermeule~\cite{vermeule2006judging} in the context of judicial decisions. In our context, errors occur not due to the erratic nature of human decision making, but because sybils do not carry an authentic representation of the society, and/or because some of the authentic voters fail to express their will.

In this sense, the addition of virtual votes in favor of the status quo should be understood as an \emph{instrumental} rather than \emph{normative} form of conservatism: it modulates the system’s responsiveness according to the level of uncertainty regarding the electorate. A bad or dysfunctional status quo can still be overturned—indeed, our liveness results formally guarantee that—once genuine participation and verified identities are sufficiently high to justify change.

Practically, this shows how institutions can remain cautious under identity uncertainty yet fully revert to the baseline social choice as legitimate participation grows.

%
}

\subsection{Related Work}\label{section:related work}

\revision{Our work complements two strands of literature: algorithms for sybil-resilient collective decision-making and economic models of institutional stability (or `resilience') under participation frictions.}
%

\paragraph{Sybils}
There is vast literature on defending against sybil attacks, see, e.g., two surveys on this topic~\cite{sybilsurvey,viswanath2010analysis}. That literature is usually concerned with graphs on which the genuine and sybil entities reside, and the focus is usually not on group decision making but on identifying the sybils. As a prominent example, Douceur~\shortcite{douceur2002sybil} describes a very general model for studying sybil resilience and presents some initial negative results in this model. Others consider leveraging graph properties such as various centrality measures to identify suspicious nodes (see, e.g.,~\cite{cao2012aiding}). As further examples, Molavi et al.~\shortcite{molavi2013iolaus} aim to shield online ranking sites from the negative effects of sybils and Chiang et al.~\shortcite{chiang2013secure} consider sybil-resilience in the context of radio networks.

We are particularly interested in sybil-resilient group decision making.
This scenario is considered by Tran et al.~\shortcite{tran2009sybil}, but with a different goal and solution:
  While we aim to protect democratic decisions from sybil attacks, they are considering ranking online content.  
Other relevant papers are the paper of Conitzer and Yokoo~\shortcite{conitzer2010using}, concentrating on axiomatic characterizations of sybil-resilient rules in a certain formal model. In essence, the authors show that in a model without a distinguished status quo alternative, the only voting rules which are sybil-safe  (no incentive for an attacker to produce sybils), is of the form ``if all vote unanimously for $c$, pick $c$, otherwise pick a winner at random''. Indeed, this negative result can also be seen as a motivation for our status quo anchored notion of sybil-safety, which enables us to provide a partial solution. 
Other relevant papers are the paper of Wagman and Conitzer~\cite{wagman2008optimal,wagman2014false}, which consider design of mechanisms to be resilient to false-name manipulation where the creation of sybils incurs some non-negligible cost. Waggoner et al.~\cite{waggoner2012evaluating} study ways to evaluate the correctness of a certain election result when the number of sybils in the electorate is assumed to be known.
Conitzer et al.~\cite{conitzer2010false} consider using connections in a social network to increase the effectiveness of sybil resilient methods.

Finally, we mention the recent  work of Gersbach et al.~\cite{gersbach2021vote,gersbach2022risky} that consider a situation with ``well-behaving'' and ``mis-behaving'' voters that share some similarities with our work, however the model is different and in particular voting is costly. \revision{We also mention the work of Lenzi~\cite{lenzi2024efficient}, who proposes an efficient mechanism that achieves sybil-attack resistance in a Bayesian setting by combining deposits and transfers; and the work of Mohan, Khezr, and Berg~\cite{mohan2024voting}, who propose \emph{bond voting}, a sybil-resistant design in which voters commit stake or time to gain influence; this contrasts with our non-monetary, rule-level approach that achieves robustness via the tunable status quo bias~$\tau$.

This line of work is complementary to ours, by lowering the amount of sybils we should expect.
}


\paragraph{Partial participation}
There is extensive work in the social choice literature on the strategic justification of partial participation/abstention, going back to the ``paradox of nonvoting"~\cite{riker1968theory,owen1984vote,desmedt2010equilibria}. 
Voting with a \emph{random set} of active voters has been widely considered, and boils down to problems of statistical estimation. See e.g. \cite{regenwetter2006behavioral,dey2015sample}. Other works consider ways to elicit the preferences of specific voters in order to reduce communication complexity~\cite{conitzer2005communication}. As far as we are aware, resilience to arbitrary partial participation has not been considered on its own, but as a special case of distorted votes (see below). 


\paragraph{`Resilience' analysis in voting}
It is quite common in the voting literature to assume that votes may deviate from the preference profile. However most of the literature assumes either some sort of stochastic noise; or strategic behavior; or both. A prominent example is the `Calculus of Voting' where voters decide whether and how to vote based on a known type distribution~\cite{riker1968theory,myerson1993theory}.

\revision{
However such distributions are rarely known either to the center or to the voters themselves, and strategic decisions may also be quite complex and relying on unknown factors. There is therefore value in understanding when results are resilient to some deviation from the benign behavior, as long as this deviation is not too large. This was done for example in the context of aggregation accuracy~\cite{procaccia2016voting}, where the authors assume that up to a certain number of bits in the profile may be corrupted, or where some of the reported preferences are distorted or omitted entirely~\cite{allouah2024robust}. The latter paper defines a property called `L-Lipschitz resilience', which can be thought of as limited voter influence on continuous outputs. Interestingly, while the authors explicitly distinguish their resilient mechanisms from SQE mechanisms that assume a status quo alternative, their proposed algorithms are implicitly biased towards a specific outcome, which effectively serves as the status quo.\footnote{The `Quadratically Regularized Median' in \cite{allouah2024robust} is biased towards  $0\in \mathbb R$, and in the `Lipschitz-Robustified Mean' the special outcome is the parameter $\mu$.} }

Two similar approaches from the side of a strategic voter include `Local Dominance', where a voter assumes the real profile may deviate from her point estimate by some margin~\cite{meir2014local}; and `Safe Manipulation', where a voter considers other similar voters may follow their behavior, but does not know how many~\cite{slinko2014ever}.

\revision{
\paragraph{Alternative motivations for conservative voting mechanisms}
Recently, Abramowitz and Mattei~\cite{abramowitz2023social} have suggested the use of conservative voting algorithms (specifically supermajority) on the grounds that they maximize \emph{worst-case acceptance} among voters, even in the absence of adversarial behavior.

While voters can try to manipulate the outcome by producing sybils, influencing the elections by the institution itself or by external lobbyists is known as \emph{control} and \emph{bribery}, respectively~\cite{controlandbribery}. 
The result of such meddling effort often manifests in adding, removing or changing some amount of votes.  We can therefore think of control and bribery attacks as yet-another-reason for the emerging of sybils and abstentions (see, e.g. \cite{colley2023measuring}). Hence, voting rules that protect against sybils provide, by extension, protection against some forms of bribery and control.
}

\subsection{Structure of the Paper}

As our model has several ingredients of differing complexity, our basic approach in structuring the paper is to start from the simplest setting, and then add the orthogonal concepts and ingredients as we go along. We feel that this allows to first grasp the basic ideas and then, as the paper proceeds, to identify the changes that are needed to be made in the model to encompass the different aspects.
Concretely, the paper is structured as follows:
\begin{itemize}

\item
In Section~\ref{section:basic} we consider the simplest social choice setting in which the voters should choose between the status quo and a single alternative proposal. We begin with the simplifying assumption of full participation and introduce the fundamental concepts of safety and liveness (Section~\ref{section:basic safety}). We then relax the assumption of full participation and formulate the adaptation of safety and liveness to this setting (Section~\ref{sec:partial}). We define the general status quo Enforcing mechanism---showing that in the simple binary setting it coincides with a supermajority rule. We analyze its safety-liveness tradeoff, and prove its optimality in Section~\ref{section:basic optimal}.

\item
In Section~\ref{sec:domains} we move beyond the binary domain of Section~\ref{section:basic},  generalizing our results for the social choice settings of multiple alternatives (Section~\ref{section:beyond ma}), multiple referenda (Section~\ref{section:beyond mr}), and to single-peaked domains (Section~\ref{section:oned}).

For this purpose, we first extend the formal definitions of safety and liveness, essentially accepting as safe outcomes that are anywhere `between' the honest outcome and the  status quo.

\item
In Section~\ref{section:relaxed} we introduce an approximate notion for safety---essentially meaning that we may arrive to an alternative that is not the preferred alternative of the honest voters, however not far from it in some sense (in particular, such that its margin of defeat is not too high). We analyze the effect of varying degrees of such approximate notion to the safety-liveness tradeoff that can be achieved for the social choice domains that are treated in Sections~\ref{section:basic} and \ref{sec:domains}.

\item
In all previous sections, the definitions of safety and liveness are worst case definitions in two aspects, considering both the behavior of sybils and the identity of absentees as adversarial, as long as their fraction is bounded. 

In Section~\ref{sec:rand_finite} we relax the latter requirement, assuming instead  that the active honest voters are chosen uniformly at random, and only requiring safety to hold \emph{with high probability} as population grows. 
%
As expected, the safety-liveness tradeoff that can be achieved in this model with a weaker adversary are better---we analyze this improved tradeoff and prove it.

\item
The final technical section (Section~\ref{sec:delegation}) considers another modification of the environment by allowing inactive voters to delegate their vote.
We show that---under common delegation assumptions---we can completely eliminate the dependency on the turnout (i.e. the \emph{fraction} of active voters),  as long as the \emph{number} of active voters is not too small.
\item
We end the paper with a discussion and an outlook (Section~\ref{section:discussion}).
 
\end{itemize}

\revision{
But first, we provide a table summarizing the main notation used throughout.
}

\revision{
\paragraph{Main notation}
To ease readability, we summarize below the key symbols used throughout the paper.
They are introduced formally as needed, but this list provides a quick reference.
}

\revision{
\begin{tcolorbox}[colback=blue!5!white,
  colframe=blue!60!black,
title=Main notation,boxrule=0.5pt,arc=2pt,]
\begin{tabular}{@{}ll@{}}
$A$ & Set of alternatives; includes the status quo $r$ and proposals $p,p',\ldots$ \\
$V$ & Set of all voters \\
$H,S$ & Sets of honest and sybil voters, respectively ($V = H \cup S$) \\
$H^+,H^-$ & Active and inactive honest voters, respectively ($H = H^+ \cup H^-$) \\
$V^+ = H^+ \cup S$ & Set of active voters (those casting a ballot) \\
$s = |S|/|V|$ & Fraction of sybil voters in the population \\
$h^+ = |H^+|/|V|$ & Fraction of active honest voters \\
$h^- = |H^-|/|V|$ & Fraction of inactive honest voters \\
$\sigma$ & Upper bound on the fraction of sybil voters ($s \le \sigma$) \\
$\mu$ & Upper bound on the fraction of inactive honest voters ($h^- \le \mu$) \\
$G$ & Base voting rule (e.g., Majority) defining the desired outcome \\
$R$ & Voting rule actually used in the election \\
$\tau$ & Fraction of virtual votes added in favor of the status quo $r$ \\
$\alpha$ & Relaxation parameter for approximate safety \\
\end{tabular}
\end{tcolorbox}
}

\section{The Basic Setting: Two Alternatives}\label{section:basic}

In this section we introduce and analyze the simplest possible setting, where there are two alternatives. The alternatives are not the same:  one of them (denoted $r$) stands for the current \emph{reality}, or \emph{status quo}, whereas the other alternative, $p$, can be viewed as a \emph{proposal} to replace it. 

Intuitively, replacing the status quo for a bad proposal is considered worse than keeping a bad status quo, but replacing the status quo should still be possible. Correspondingly, we define the basic concepts of safety and liveness in face of sybils and partial participation, and show how the best trade off between them can be obtained in the worst case. 


\subsection{Preliminaries}


We consider voting situations with a set $A = \{r, p\}$ of  alternatives, with $r$ referred to as the current \emph{reality}, or \emph{status quo} and $p$ is a competing \emph{proposition/proposal}.

There is a set $V$ of $n$ voters, each specifying whether she prefers $r$ to $p$ or vice versa. A voting rule is a function taking the $n$ votes and returning an outcome in $A$. Most social choice settings we consider in this paper are such that each voter votes by picking an alternative and the aggregated outcome is also an alternative; thus, a voting rule $\calR$ is a sequence of functions $\calR^n : A^n \rightarrow A$, for all $n\in \mathbb N$.




\mypara{Honest voters and sybils}

The set of voters $V$ is partitioned into a set of \emph{honest} (i.e., genuine; non-sybil) voters~$H$ and a set of \emph{sybil} voters $S$; so, $V = H \cup S$ with $H \cap S = \emptyset$. We assume there is always at least one honest voter, so $H\neq \emptyset$. Ideally, we would like our voting rules to reflect only the preferences of the honest voters, but without access to who is honest and who is sybil, and when not all honest voters vote.

\begin{figure}
    \centering

\begin{tikzpicture}[scale=0.45]

\tikzstyle{dot}=[rectangle,draw=black,fill=white,inner sep=0pt,minimum size=4mm]
\tikzstyle{active}=[circle,draw=blue,text=white,fill=blue,inner sep=0pt,minimum size=3mm]
\tikzstyle{inactive}=[circle,draw=blue,fill=white,inner sep=0pt,minimum size=3mm]
\tikzstyle{sybil}=[rectangle,draw=black,fill=red,text=white,inner sep=0pt,minimum size=3mm]
\tikzstyle{del}=[triangle,draw=blue,fill=white,inner sep=0pt,minimum size=2mm]
\tikzstyle{virtual}=[diamond,draw=black,fill=gray,inner sep=0pt,minimum size=4mm]
\tikzstyle{txt}=[text width = 2cm, anchor=west]
\node at (-2,0) {};

\draw (-2,0.4) -- (8,0.4);
\node at (0,-.2) {$r$};
\node at (5,-.2) {$p$};

\node at (0,1) [inactive] {};

\node at (-0.5,2) [active] {};
\node at (+0.5,2) [active] {};

\node at (5-1,1) [inactive] {};
\node at (5,1) [inactive] {};
\node at (5+1,1) [inactive] {};

\node at (5,2) [active] {};

\node at (5-0.5,3) [sybil] {};
\node at (5+0.5,3) [sybil] {};



\node at (-2.5, 1) {$H^-$};
\node at (-2.5, 2) {$H^+$};
\node at (-2.5, 3) {$S$};
\draw [decorate,decoration={brace,amplitude=5pt},xshift=-0pt,yshift=-4pt]
(-3.5,0.5) -- (-3.5,2.5) node [black,midway,xshift=-1.8cm] 
{Honest voters $H$};
\draw [decorate,decoration={brace,amplitude=10pt},xshift=-0pt,yshift=-4pt]
(8,3.8) -- (8,1.8) node [black,midway,xshift=+2cm] 
{Active voters $V^+$};

\end{tikzpicture}
    \caption{\label{fig:vhs}
    Example of a voting setting with two alternatives $A=\{r,p\}$. There are $|V|=9$ voters overall, of which $|S|=2$ are sybils, and $|H^-|=4$ are inactive. Therefore $s=\frac{2}{9}$ and $h=\frac{7}{9}$.  Similarly, $h_p = \frac{|H_p|}{|V|}=\frac{4}{9}$ as there are 4 honest voters for $p$. 
    We keep using full/hollow blue circles for active / inactive voters and red squares for sybils throughout the paper.}
\end{figure}

\mypara{Further notation}
In many places it will be convenient to refer to the \emph{fraction} of some set of voters rather than to their absolute size. For any subset of voters $U\subseteq V$, we denote by the lowercase letter $u:=\frac{|U|}{|V|}$ the relative size of this set to the entire population. 

We denote by $U_a\subseteq U$ the subset of $U$ voters who prefer alternative $a\in A$, and by $u_a=\frac{|U_a|}{|V|}$ their relative fraction. An example of the different voters' types and their notation is in Fig.~\ref{fig:vhs} \revision{; refer also to the main notation table at the end of Section~\ref{section:introduction}}.

Crucially, we do not know up front how many voters are sybils. However, we assume for the purpose of analysis that the fraction of sybils voters---which we denote by $s:=\frac{|S|}{|V|}$---is upper bounded by the known parameter $\sigma\in[0,1)$ (in Section~\ref{section:discussion} we discuss how to estimate such value). Thus, higher values of $\sigma$ allow for a wider range of instances, and more difficult ones.

\subsect{Safety and Liveness (Full Participation)}\label{section:basic safety}


Suppose we have some preferred voting rule $\calG$, for the ``standard'' setting without sybils and with full participation. This may be due to favorable axiomatic or social properties of $\calG$, because of its simplicity, due to legacy, or for any other reason. For the setting of two alternatives, the Majority rule is natural -- see discussion below as well. Ideally, we would like to always get outcome $\calG(H)$, that is, the result of all honest voters voting under $\calG$. However, if we use $\calG$ in a straightforward way, then the outcome may be distorted due to the existence of sybil votes, due to the partial participation, or both. 

\begin{example}[Sybils]\label{ex:sybils}
As a simple example, say that we would like to use Majority rule $\MJ$. Our population consists of five active honest voters, three of which vote for $r$ and two for $p$. 
However if we add two sybils voting for $p$ then the less desired outcome $p$ now has most votes; see left figure.
\begin{center}
\begin{tikzpicture}[scale=0.4]

\tikzstyle{dot}=[rectangle,draw=black,fill=white,inner sep=0pt,minimum size=4mm]
\tikzstyle{active}=[circle,draw=blue,text=white,fill=blue,inner sep=0pt,minimum size=3mm]
\tikzstyle{inactive}=[circle,draw=blue,fill=white,inner sep=0pt,minimum size=3mm]
\tikzstyle{sybil}=[rectangle,draw=black,fill=red,text=white,inner sep=0pt,minimum size=3mm]
\tikzstyle{del}=[triangle,draw=blue,fill=white,inner sep=0pt,minimum size=2mm]
\tikzstyle{virtual}=[diamond,draw=black,fill=gray,inner sep=0pt,minimum size=4mm]
\tikzstyle{txt}=[text width = 2cm, anchor=west]

\node at (3.5,4) {\textit{unsafe}};

\draw (-1,0.4) -- (8,0.4);
\node at (1,-.2) {$r$};
\node at (6,-.2) {$p$};

\node at (0,2) [active] {};
\node at (1,2) [active] {};
\node at (2,2) [active] {};

\node at (6-0.5,2) [active] {};
\node at (6+0.5,2) [active] {};


\node at (6-0.5,3) [sybil] {};
\node at (6+0.5,3) [sybil] {};

\end{tikzpicture}
~~~~~~~~~~~~~~~~ 
\begin{tikzpicture}[scale=0.4]

\tikzstyle{dot}=[rectangle,draw=black,fill=white,inner sep=0pt,minimum size=4mm]
\tikzstyle{active}=[circle,draw=blue,text=white,fill=blue,inner sep=0pt,minimum size=3mm]
\tikzstyle{inactive}=[circle,draw=blue,fill=white,inner sep=0pt,minimum size=3mm]
\tikzstyle{sybil}=[rectangle,draw=black,fill=red,text=white,inner sep=0pt,minimum size=3mm]
\tikzstyle{del}=[triangle,draw=blue,fill=white,inner sep=0pt,minimum size=2mm]
\tikzstyle{virtual}=[diamond,draw=black,fill=gray,inner sep=0pt,minimum size=4mm]
\tikzstyle{txt}=[text width = 2cm, anchor=west]

\node at (3.5,4) {\textit{safe}};

\draw (-1,0.4) -- (8,0.4);
\node at (1,-.2) {$r$};
\node at (6,-.2) {$p$};

\node at (6-1,2) [active] {};
\node at (6,2) [active] {};
\node at (6+1,2) [active] {};

\node at (1-0.5,2) [active] {};
\node at (1+0.5,2) [active] {};


\node at (1-0.5,3) [sybil] {};
\node at (1+0.5,3) [sybil] {};

\end{tikzpicture}
    \end{center}
 Intuitively, this means Majority is \emph{unsafe} in the presence of sybils. 
 
We can also think about the opposite situation where the honest voters want $p$, but the Majority rule maintains the status quo due to the presence of sybils, as in the right figure. 
This is \emph{not} considered a violation of safety, since maintaining the status quo is always safe.
\end{example}

\paragraph{Base rules and the Majority rule}

Note that the Majority rule plays a double role in our examples above: it defines what is the desired outcome, and is also the voting rule being used.

In general we may use different rules for these roles: we will denote the \emph{base rule} (which sets the desired outcome) by $\calG$. Throughout this section, the base rule $\calG$ will always be the Majority rule $\MJ$, as this is the only rule that is monotone, anonymous, and neutral~\cite{may1952set}, and is generally the rule that makes most sense. We will use $\calR$ to denote the rule used in practice. Also, unless explicitly stated otherwise, we always assume $\MJ$ breaks ties in favor of $r$.



\revision{
\paragraph{Safety}
The previous examples show that the simple Majority rule can be \emph{unsafe}: it may trigger an undesired change from the status quo when sybils are present. 

We now formalize this notion for the two-alternative setting.  
To keep the exposition simple, we begin with the case of full participation and later extend it to partial participation in Section~\ref{sec:partial}.

\begin{definition}[Safety, two alternatives, full participation]\label{def:safe}
A voting rule $\mathcal{R}$ is \emph{safe} with respect to a base rule $\mathcal{G}$ and an active population $V = H \cup S$ if
\[
\mathcal{R}(V) \in \{ \mathcal{G}(H),\, r \}.
\]
\end{definition}

\noindent
In words, a rule is safe when it never selects an outcome that honest voters would reject. 
It either reproduces the result that the base rule $\mathcal{G}$ would have produced using only honest votes, or it leaves the status quo $r$ unchanged.

\medskip
\noindent
For example, the Majority rule is \emph{not} safe with respect to itself for the population in Example~\ref{ex:sybils}, since $\MJ(V) = p$ whereas $\{\MJ(H), r\} = \{r\}$.
}

\revision{
\paragraph{Liveness}
If safety were our only requirement, achieving it would be trivial: one could simply keep the status quo $r$ regardless of how people vote. 
However, such a rule would render participation meaningless. 
We therefore impose a complementary requirement: that the honest population must be able to \emph{enforce any desired outcome} through voting.
Liveness captures this property. 
Unlike safety, it does not rely on a base rule or on any specific structure of alternatives.

\medskip
\noindent
For an outcome $a \in A$ and a set of votes $U$, let $U_{\rightarrow a}$ denote the same population $U$ in which all voters vote for $a$ (while their types remain unchanged).

\begin{definition}[Liveness, full participation]\label{def:live}
A voting rule $\mathcal{R}$ is \emph{live} with respect to an active population $V = H \cup S$ if, for every alternative $a \in A$,
\[
\mathcal{R}(S \cup H_{\rightarrow a}) = a.
\]
\end{definition}

\noindent
In words, a rule is live when honest voters can always realize any outcome they unanimously support, regardless of how sybils vote.

\medskip
\noindent
Without further restrictions, safety and liveness may be incompatible—for instance, when almost the entire population consists of sybils. 
We therefore study which rules can satisfy both properties under bounded fractions of problematic voters.
For example, the \emph{Unanimity} rule is safe whenever $\sigma < 1$ but fails liveness for any $\sigma > 0$.
A rule that requires a supermajority of $3/4$ in favor of $p$ and otherwise keeps $r$ is safe (under full participation) for $\sigma = \tfrac{1}{3}$ but not for $\sigma = \tfrac{2}{3}$.
Our goal is to understand the best attainable trade-off between safety and liveness.
}

\subsect{Partial Participation}\label{sec:partial}

Next, we introduce to the model the possibility of voters to abstain from the vote.
We need some further notation and definitions to capture this aspect first.

\mypara{Active and passive voters}
Recall that the set of voters $V$ is partitioned into a set of honest voters $H$ and a set of sybil voters $S$. As we assume the worst case, w.l.o.g.  all sybil voters participate; but the set of honest voters, $H$, is further partitioned into $H = H^+ \cup H^-$ (with $H^+ \cap H^- = \emptyset$), where $H^+$ is the non-empty set of honest voters who did cast a vote, and are thus labeled by their vote, and $H^-$ is the set of honest voters who did not cast a vote. We refer to the voters in $H^+$ as \emph{active honest voters} and to the voters in $H^-$ as \emph{passive honest voters}, or \emph{passive voters} in short.  
\revision{
Thus, in the partial-participation setting, both the active honest voters and the sybils cast ballots. 
We denote the set of all active voters by
\[
V^{+} := H^{+} \cup S,
\]
and observe that the entire population can now be written as
\[
V = H^{+} \cup H^{-} \cup S.
\]
In words, $H^{+}$ are the honest voters who participate, $H^{-}$ are those who abstain, and $S$ are the sybils.
}



\paragraph{Further notation}
As with the rate of sybils, we do not know up front how many voters are inactive. However, we assume for the purpose of analysis that the fraction of inactive voters---which we denote by $h^-:=\frac{|H^-|}{|V|}$---is upper bounded by the known parameter $\mu\in[0,1)$ (we discuss in Section~\ref{section:discussion} how to estimate this value). Thus, higher value of $\mu$ allows for a wider range of instances, and more difficult ones.

\begin{example}[Abstention]\label{ex:abstain}
Consider an example without sybils, where $|H_r|=3$ and $|H_p|=2$. 
If two of the active voters for $r$ abstain (i.e. $|H^+_r|=1$ then $p$ would win, see left figure: 

\begin{center}
\begin{tikzpicture}[scale=0.4]

\tikzstyle{dot}=[rectangle,draw=black,fill=white,inner sep=0pt,minimum size=4mm]
\tikzstyle{active}=[circle,draw=blue,text=white,fill=blue,inner sep=0pt,minimum size=3mm]
\tikzstyle{inactive}=[circle,draw=blue,fill=white,inner sep=0pt,minimum size=3mm]
\tikzstyle{sybil}=[rectangle,draw=black,fill=red,text=white,inner sep=0pt,minimum size=3mm]
\tikzstyle{del}=[triangle,draw=blue,fill=white,inner sep=0pt,minimum size=2mm]
\tikzstyle{virtual}=[diamond,draw=black,fill=gray,inner sep=0pt,minimum size=4mm]
\tikzstyle{txt}=[text width = 2cm, anchor=west]

\node at (3.5,4) {\textit{unsafe}};

\draw (-1,0.4) -- (8,0.4);
\node at (1,-.2) {$r$};
\node at (6,-.2) {$p$};

\node at (0.5,1) [inactive] {};
\node at (1,2) [active] {};
\node at (1.5,1) [inactive] {};

\node at (6-0.5,2) [active] {};
\node at (6+0.5,2) [active] {};



\end{tikzpicture}
~~~~~~~~~~~~~~    
\begin{tikzpicture}[scale=0.4]

\tikzstyle{dot}=[rectangle,draw=black,fill=white,inner sep=0pt,minimum size=4mm]
\tikzstyle{active}=[circle,draw=blue,text=white,fill=blue,inner sep=0pt,minimum size=3mm]
\tikzstyle{inactive}=[circle,draw=blue,fill=white,inner sep=0pt,minimum size=3mm]
\tikzstyle{sybil}=[rectangle,draw=black,fill=red,text=white,inner sep=0pt,minimum size=3mm]
\tikzstyle{del}=[triangle,draw=blue,fill=white,inner sep=0pt,minimum size=2mm]
\tikzstyle{virtual}=[diamond,draw=black,fill=gray,inner sep=0pt,minimum size=4mm]
\tikzstyle{txt}=[text width = 2cm, anchor=west]

\node at (3.5,4) {\textit{safe}};

\draw (-1,0.4) -- (8,0.4);
\node at (1,-.2) {$r$};
\node at (6,-.2) {$p$};

\node at (6-0.5,1) [inactive] {};
\node at (6,2) [active] {};
\node at (6+0.5,1) [inactive] {};

\node at (1-0.5,2) [active] {};
\node at (1+0.5,2) [active] {};



\end{tikzpicture}
    \end{center}
    This again demonstrates how abstention, just like sybil participation, may lead to an unsafe outcome. As in Example~\ref{ex:sybils}, if abstention results in the selection of $r$ (as in the right figure), we should not consider this a violation of safety.


\end{example}

\revision{
\paragraph{Safety and Liveness under Partial Participation}
We now extend the previous notions of safety and liveness to settings where not all honest voters participate. 
Let $V^{+} = H^{+} \cup S$ denote the set of active voters (honest and sybil), and recall that the full population is $V = H^{+} \cup H^{-} \cup S$.

In turning this intuition into a formal definition, note that the desired outcome $\mathcal{G}(H)$ is determined exactly as in Definition~\ref{def:safe}, but the realized outcome is computed only on the active voters.
Since both sybils and inactive honest voters are indistinguishable from the perspective of the voting rule, we assume that no mechanism can tell members of $S$ apart from those in~$H^{+}$.

\begin{definition}[Safety and Liveness under Partial Participation]\label{def:safe_live_partial}
For a voting rule $\mathcal{R}$ and base rule $\mathcal{G}$:
\begin{itemize}
    \item $\mathcal{R}$ is \emph{safe} with respect to $\mathcal{G}$ and $V = H \cup S$ if
    \[
    \mathcal{R}(V^{+}) \in \{ \mathcal{G}(H),\, r \}.
    \]
    \item $\mathcal{R}$ is \emph{live} with respect to $V = H \cup S$ if, for every $a \in A$,
    \[
    \mathcal{R}(S \cup H^{+}_{\rightarrow a}) = a.
    \]
\end{itemize}
\end{definition}
Note that the only difference from Definitions \ref{def:safe} and \ref{def:live} above is the emphasis that $\calR$ operates only on the active votes. 

}



We already saw that the Majority rule is not safe with respect to itself even if all voters are active (Example~\ref{ex:sybils}) or if there are no sybils (Example~\ref{ex:abstain}).

\paragraph{Example under Supermajority}
Consider the $3/4$-supermajority rule $\calR'$. We argue that, for the left instance in Example~\ref{ex:sybils}, $\calR'$ is both safe and live. To see why it is safe, note first that $\calR'(V^+)$ also selects $r$, since $p$ only has a $4/7$-majority which is less than $3/4$. To see why it is live, note that if all honest voters switch to $p$, then $S\cup H^+_{\rightarrow p}$ has 7 votes to $p$ vs. 0 votes to $r$, so the $3/4$-supermajority rule will select $p$. 

It is also not hard to see that $\calR'$ is safe and live for both instances in Example~\ref{ex:abstain}.
 In contrast, in the right instance in Example~\ref{ex:sybils},  there is nothing honest voters can do to get $p$ selected under  rule $\calR'$. Therefore $\calR'$ is not live for that instance. Weakening the supermajority requirement to anything strictly below $4/7$ would regain liveness, since it would enable the three active honest voters to obtain any outcome.  Clearly for every instance there is \emph{some} supermajority threshold above which safety is guaranteed, and likewise, some threshold below which liveness is guaranteed.

\subsect{Optimal Safety-Liveness Tradeoff}\label{section:basic optimal}

\revision{
\paragraph{Supermajority as a compromise between safety and liveness}
The examples above suggest that a practical way to achieve both safety and liveness is to relax neutrality and give a slight formal preference to the status quo~$r$. 
Intuitively, we allow change to a proposal~$p$ only when it is supported by a sufficiently large majority—large enough to protect against sybils, but not so large as to paralyze collective choice.

\begin{definition}[Supermajority rule]\label{def:SMJ}
Let $A = \{ r, p \}$. 
For a threshold parameter $\tau \in [0,1)$, the \emph{$\tau$-Supermajority rule} ($\tau$-SM) selects $p$ if the fraction of votes for $p$,
denoted $v_p$, satisfies
\[
v_p > \frac{1+\tau}{2},
\]
and selects $r$ otherwise. In particular, ties are broken in favor of $r$.
\end{definition}

\noindent
In words, the rule approves the proposal only when its support exceeds a $(1+\tau)/2$ fraction of the votes, thus requiring a stronger mandate for change as $\tau$ increases.
}

Indeed, if we restrict attention to anonymous and monotone rules, then there is not much else we could do. Intuitively, as we increase the supermajority we require, we get more safety (i.e., for higher rates of sybils and abstention), but less liveness.\footnote{Here `more' refers to the range of instances on which safety or liveness can be obtained. Later in Section~\ref{section:relaxed} we propose an additional way to quantify safety  of a rule on a given instance.}  

Our goal is to characterize this tradeoff.

\revision{
\paragraph{The status quo–Enforcing mechanism}
A second way to balance safety and liveness is to modify the majority rule by introducing \emph{virtual voters} who always support the status quo~$r$. 
Informally, the \emph{status quo–Enforcing} (SQ) mechanism adds a fixed fraction of such virtual votes to the electorate before applying the underlying rule.

\begin{definition}[status quo–Enforcing mechanism]\label{def:RE}
Let $\mathcal{R}$ be a voting rule. 
For a parameter $\tau \ge 0$, define the mechanism
\[
\tau\text{-}\mathrm{SQ}\text{-}\mathcal{R}(V) := \mathcal{R}(V^+ \cup Q),
\]
where $Q$ is a set of $\tau |V^{+}|$ virtual voters who all vote for the status quo~$r$.\footnote{The quantity $\tau |V^{+}|$ may be fractional, but for most rules—including Majority—this poses no difficulty.} Note that if there are inactive votes in $V$, the rule is simply unaware of those.
\end{definition}

\noindent
In words, the $\tau$-SQ-$\mathcal{R}$ mechanism behaves as if an additional $\tau$-fraction of the active electorate were voting for $r$. 
The larger $\tau$ is, the more conservative the rule becomes, since it requires stronger support for proposals to overturn the status quo.

}

\revision{As explained in the introduction, the added virtual voters are not intended to emphasize a normative advantage of the status quo, but rather to protect against unverified and uncertain change.}

\revision{
\paragraph{Relationship to supermajority}
The status quo–Enforcing mechanism can be instantiated with the Majority rule, in which case it produces exactly the supermajority rule. 
Intuitively, adding $\tau$ virtual votes for the status quo requires the proposal $p$ to achieve a $(1+\tau)/2$ fraction of support before being accepted.

\begin{observation}
For $A = \{ r, p \}$ and any $\tau \ge 0$, the $\tSQMJ$ rule and the $\tau$-SM rule coincide.
\end{observation}

\full{
\begin{proof}
Let $v_p:=\frac{|V_p|}{|V|}$ denote the fraction of votes for $p$. 
The $\tSQMJ$ rule selects $p$ when
\[
v_p > v_r + q_r = (1 - v_p) + \tau,
\]
which is equivalent to
\[
v_p > \frac{1+\tau}{2}.
\]
Hence, $\tSQMJ$ and $\tau$-SM yield identical outcomes.
\end{proof}
}
}

\begin{figure}[t]
    \centering
\begin{tikzpicture}[scale=0.4]

\tikzstyle{dot}=[rectangle,draw=black,fill=white,inner sep=0pt,minimum size=4mm]
\tikzstyle{active}=[circle,draw=blue,text=white,fill=blue,inner sep=0pt,minimum size=3mm]
\tikzstyle{inactive}=[circle,draw=blue,fill=white,inner sep=0pt,minimum size=3mm]
\tikzstyle{sybil}=[rectangle,draw=black,fill=red,text=white,inner sep=0pt,minimum size=3mm]
\tikzstyle{del}=[triangle,draw=blue,fill=white,inner sep=0pt,minimum size=2mm]
\tikzstyle{virtual}=[diamond,draw=black,fill=gray,inner sep=0pt,minimum size=3.2mm]
\tikzstyle{txt}=[text width = 2cm, anchor=west]

\draw (-1,0.4) -- (7,0.4);
\node at (1,-.2) {$r$};
\node at (6,-.2) {$p$};

\node at (-4,4) {virtual votes};

\node at (1-1,2) [active] {};
\node at (1+0,2) [active] {};
\node at (1+1,2) [active] {};

\node at (6-0.5,2) [active] {};
\node at (6+0.5,2) [active] {};

\node at (1-0.5,4) [virtual] {};
\node at (1+0.5,4) [virtual] {};


\node at (6-0.5,3) [sybil] {};
\node at (6+0.5,3) [sybil] {};

\end{tikzpicture}
~~~~~~~~~~~~~~~~~~~
\begin{tikzpicture}[scale=0.4]

\tikzstyle{dot}=[rectangle,draw=black,fill=white,inner sep=0pt,minimum size=4mm]
\tikzstyle{active}=[circle,draw=blue,text=white,fill=blue,inner sep=0pt,minimum size=3mm]
\tikzstyle{inactive}=[circle,draw=blue,fill=white,inner sep=0pt,minimum size=3mm]
\tikzstyle{sybil}=[rectangle,draw=black,fill=red,text=white,inner sep=0pt,minimum size=3mm]
\tikzstyle{del}=[triangle,draw=blue,fill=white,inner sep=0pt,minimum size=2mm]
\tikzstyle{virtual}=[diamond,draw=black,fill=gray,inner sep=0pt,minimum size=3.2mm]
\tikzstyle{txt}=[text width = 2cm, anchor=west]

\draw (-1,0.4) -- (7,0.4);
\node at (1,-.2) {$r$};
\node at (6,-.2) {$p$};

\node at (1-0.5,1) [inactive] {};
\node at (1+0,2) [active] {};
\node at (1+0.5,1) [inactive] {};

\node at (6-0.5,2) [active] {};
\node at (6+0.5,2) [active] {};

\node at (1,4) [virtual] {};



\end{tikzpicture}
    \caption{\label{fig:running_ex_sq} Two instances from the previous examples, where Majority is unsafe with respect to itself but adding  virtual voters (gray diamonds) restores safety with respect to Majority.}
\end{figure}

The reason why we will be focusing on $\tSQMJ$ rather than on $\tSM$ is that the former naturally generalizes to other domains (see Section~\ref{sec:domains}).

\paragraph{Safety and Liveness of the status quo Enforcing Mechanism} 
The $\tSQMJ$ is safer than Majority. We can see that on our running examples.
 \begin{example}[status quo Enforcing Majority]
 Consider the unsafe instance from Example~\ref{ex:sybils}. Applying $\frac27\shyp\REMJ$ to this instance would add $\frac27|V^+|=2$ virtual voters on $r$ (see Fig.~\ref{fig:running_ex_sq}, Left). Thus $\frac27\shyp\REMJ(V)=\MJ(V\cup Q)$ with $5$ voters on $r$ vs. only $4$ on $p$. So $r\in \{\MJ(H),r\}$ wins and safety is restored. 

 Similarly, applying $\frac13\shyp\REMJ$ to the unsafe instance from Example~\ref{ex:abstain} adds $\frac13|V^+|=1$ virtual voter on $r$ (see Fig.~\ref{fig:running_ex_sq}, Right). By tie-breaking $r$ wins so there is no violation of safety. 
 \end{example}
 
\revision{
\paragraph{Safety–liveness trade-off in the binary setting}
We are now ready to address the main question: under what conditions can both safety and liveness be achieved simultaneously? 
The proof of safety is omitted here, as it follows as a special case of the more general Theorem~\ref{thm:SQMJ_safe_alpha} presented later.

\begin{theorem}\label{thm:SQMJ}
For the binary setting $A = \{ r, p \}$, the following hold:
\begin{itemize}
    \item $\tSQMJ$ is \emph{safe} with respect to Majority if and only if
    \[
    \tau \ge 
    \frac{\sigma+\mu}{1-\mu}.
    \]
    \item $\tSQMJ$ is \emph{live} if and only if
    \[
    \tau < 
    \frac{1-2\sigma-\mu}{1-\mu}.
    \]
\end{itemize}
That is, the valid interval in which both safety and liveness are guaranteed is 
\begin{center}\boxed{\frac{\sigma+\mu}{1-\mu} \leq \tau < \frac{1-2\sigma-\mu}{1-\mu}~~~~\textit{  iff  }~~~~~3\sigma + 2\mu < 1}\end{center}

\end{theorem}

\noindent
In words, the parameter $\tau$ determines how conservative the system is: larger values protect better against sybils and abstentions (safety), while smaller values allow change more easily (liveness). 
The two inequalities specify exactly when both goals can be met. 
For example, with $20\%$ sybils and $20\%$ abstentions, or with $10\%$ sybils and $35\%$ abstentions, both properties hold as long as the above condition is satisfied.
}

One way to visualize the safety-liveness tradeoff is in Fig.~\ref{fig:tradeoff}. We can see that when $\sigma$ and $\mu$ are low (meaning few sybils and low abstention), there is a wide range of mechanisms that are both safe and live, but this range diminishes as $\sigma$ and/or $\mu$ is increasing, becoming empty when $3\sigma+2\mu\geq 1$.

\revision{
Before turning to the formal argument, recall that liveness means that whenever all honest voters coordinate on supporting the proposal~$p$, the mechanism must indeed select~$p$. 
The challenge is that sybils and inactive honest voters effectively add weight to the status quo~$r$, while the parameter~$\tau$ further amplifies this bias through virtual votes. 
The proof below verifies that, as long as $\tau$ is smaller than the stated threshold, the honest support for~$p$ dominates these opposing forces.
}

\begin{proof}[Proof of Theorem~\ref{thm:SQMJ} (liveness)]
    Suppose first that $\tau< \frac{1-2\sigma-\mu}{1-\mu}$. The worst case for liveness is when all voters are on $r$.

In the profile $H_{\rightarrow p}$ there will be $|H^+|\geq (1-\mu-\sigma)|V|$ active votes for $p$, vs. at most $\sigma|V|+\tau|V^+|=(\sigma+\tau(1-\mu))|V|$ active votes for $r$. We compare:
\begin{align*}
     v^+_r &\leq \sigma+\tau(1-\mu) <  \sigma+\left(\frac{1-2\sigma-\mu}{1-\mu}\right)(1-\mu) \\
     &= \sigma+1-2\sigma-\mu  = 1 - \sigma -\mu \leq v^+_p,
\end{align*}
so $p$ is selected.

In the other direction, set $s=\sigma$ and $h^-=\mu$ and then all weak inequalities become equalities, and the strict inequality flips, so $\MJ(H_{\rightarrow p})=r$.
\end{proof}

\begin{figure}
    \centering

\begin{tikzpicture}
  \begin{axis}[
    xmin=0, xmax=1,
    ymin=0, ymax=2.3,
    axis lines=middle,
    xlabel={ abstention $\mu$},
        x label style={at={(axis description cs:0.5,-0.18)},anchor=south},
    ylabel={parameter $\tau$ (virtual votes)},
    y label style={at={(axis description cs:-0.1,.5)},rotate=90,anchor=south},
    xtick={0.2,0.35,0.6,0.8},
    ytick={0,0.1,0.6,0.8,1,2},
    legend style={at={(axis cs:1,1.5)},anchor=north,legend cell align=left}
    ]
    
    \addplot[blue, thick, domain=0:1-1.1/3, samples=100, name path=f_safe] {(1+0.1)/(1-x)-1};
      \addlegendentry{safety ($\sigma=0.1$)}
    \path[name path=Haxis] (axis cs:0,2) -- (axis cs:1,2);
    \addplot[blue!30,semitransparent,forget plot] fill between[of=f_safe and Haxis, soft clip={domain=0:1}];
    \addplot[red, thick, domain=0:0.99, samples=100, name path=f_live] {2*(1-0.1-x)/(1-x)-1};
          \addlegendentry{liveness ($\sigma=0.1$)}

    \path[name path=Laxis] (axis cs:0,0) -- (axis cs:1,0);
    \addplot[red!30,semitransparent,forget plot] fill between[of=f_live and Laxis, soft clip={domain=0:1}];
  \node[blue] at (axis cs: 0.25,1.2) {\LARGE{safe}};
    \node[red] at (axis cs: 0.50,0.35) {\LARGE{live}};
        \addplot[blue, thick, dotted, domain=0:1-1.2/3, samples=100] {(1+0.2)/(1-x)-1};
              \addlegendentry{safety ($\sigma=0.2$)}

    \addplot[red, thick, dotted, domain=0:0.99, samples=100] {2*(1-0.2-x)/(1-x)-1};
                  \addlegendentry{liveness ($\sigma=0.2$)}

    \draw[dashed,gray] (axis cs:0.2,0) -- (axis cs:0.2,0.5);
        \draw[dashed,gray] (axis cs:0.35,0) -- (axis cs:0.35,9/13);

  \end{axis}
\end{tikzpicture}
\caption{In this figure (solid lines) the fraction of sybils is fixed at $\sigma=0.1$, i.e. 10\% sybils. For every value of abstention $\mu$, we color in blue the range of $\tSQMJ$ mechanisms that are safe. The range of live mechanisms is in red.  The dotted lines mark the ranges when there are 20\% sybils rather than 10\%. \revision{Recall that $\tau$ denotes the fraction of virtual votes added for the status quo~$r$, $\sigma$ is the fraction of sybils, and $\mu$ is the fraction of inactive honest voters.}
}

    \label{fig:tradeoff}
\end{figure}

\revision{
\paragraph{Lower bound}
We complement our analysis with a lower bound that establishes the tightness of the previous result. 
Intuitively, when the combined fraction of sybils and abstaining honest voters exceeds the critical threshold $3\sigma + 2\mu = 1$, 
no rule can simultaneously guarantee both safety and liveness: any attempt to remain responsive makes the system vulnerable to manipulation, 
while any rule that resists manipulation becomes permanently inert.
}

\begin{theorem}\label{thm:arb_LB}
There is no mechanism $\calR$ such that  $\calR$ is both safe (with respect to Majority) and live when $3\sigma+2\mu\geq 1$.\footnote{
We assume that there is at least one honest voter, otherwise safety is meaningless.}
\end{theorem}


\begin{proof}
Assume towards a contradiction that such a mechanism $\calR$ exists. 
By liveness, there is a profile $V$ with $s_r=\sigma$ (i.e. all allowed sybils exist and are voting for $r$), and yet $p$ is selected, i.e. $\calR(V^+)=\calR(S\cup H^+)=p$. The total number of active voters for $p$ is $h^+_p$. Note that $h^+_p\leq h^+ \leq 1-\mu-\sigma$.

Now, consider a profile $\ol V=\ol S\cup \ol H^+ \cup \ol H^-$, where $|\ol S|=|S|, |\ol H^+|=|H^+|, |\ol H^-|=|H^-|$, so $\sigma$ and $\mu$ are still respected in $\ol V$. Set $\ol s_p:=\min\{h^+_p,\sigma\}$ sybils to vote for $p$, as well as exactly $\ol h^+_p:=h^+_p-\ol s_p$ honest voters. All other voters vote for $r$ (including all inactive honest voters). Since $\ol v^+_p=v^+_p$ and $\ol v^+_r=v^+_r$, the profiles $V$ and $\ol V$ are indistinguishable for $\calR^+$, and we have $\calR^+(\ol V)=p$ as well.

We will show that $\ol h_r\geq \ol h_p$, which entails a violation of safety. Suppose first that $\sigma<h^+_p$.
Then, 
\begin{align*}
    \ol h_r -\ol h_p &= (\ol h^- + \ol h^+_r) - \ol h^+_p = (\ol h^- + \ol h^+ - \ol h^+_p) - \ol h_p \\
    & = \ol h^- + \ol h^+ - 2\ol h^+_p = \mu +(1-\sigma-\mu) - 2\ol h^+_p \\
    &=1-\sigma - 2(h^+_p-\ol s_p) = 1-\sigma - 2(h^+_p-\sigma)\\
    &= 1+\sigma -2h^+_p  \geq  1+\sigma -2(1-\mu-\sigma)\\
    & = 3\sigma + 2\mu -1 \geq 0,
\end{align*}
where the last inequality is by the premise of the theorem. 

If $\sigma\geq h^+_p$, then 
$$\ol h_p^+=h_p^+-\min\{h^+_p,\sigma\}= h_p^+-h^+_p = 0,$$ 
i.e., $\ol V$ contains no honest voters for $p$ at all, which means $\ol h_r > \ol h_p$.
\end{proof}


\section{Beyond the Binary Domain}\label{sec:domains}
The modification we applied to the Majority voting rule simply added `virtual votes' on the status quo. It is not hard to see that this idea easily extends to many other domains, i.e. that $\tRER$ is well-defined for any voting rule $\calR$ in any domain were votes can be thought of as positions in some space.  

However, our current definition of safety is too narrow. For example, suppose that $A$ is the real line, the status quo is $r=0$, and some rule $\calG$ is our base rule (say, Median). If the honest population prefers $\calG(H)=3$, then only  '$0$' and '$3$' are considered `safe'. But if we are willing to accept both '$0$' and '$3$', then it makes sense to all accept all outcomes in between.  Indeed this is the logic behind our general definition of \emph{between set} below.

Our definition of liveness also needs an adaptation: The space of allowed ballots may not coincide with $A$, and thus $H_{\rightarrow a}$ may not be well-defined.

\def\CC{\textit{CC}}
\def\SCC{\textit{SCC}}

%

\revision{
\paragraph{The notion of betweenness}
To reason about more general domains of alternatives, we model the alternative set~$A$ as a metric space~$(A,\delta)$, 
where $\delta$ measures the distance between alternatives. 
Each vote specifies one alternative in this space, and the distance function allows us to formalize notions such as “closeness” or “compromise” between alternatives.

\medskip
\noindent
Every metric space $(A,\delta)$ naturally induces a ternary relation of \emph{betweenness}: 
an alternative $b$ is said to be \emph{between} $a$ and $c$ if
\[
\delta(a,b) + \delta(b,c) = \delta(a,c),
\]
as in classical geometric formulations of betweenness~\cite{menger1928untersuchungen,chvatal2004sylvester}.

\begin{definition}[Between set]\label{definition:between}
For $x, y \in A$, let $\mathcal{B}(x,y) \subseteq A$ be the set of all points that are between $x$ and $y$, including $x$ and $y$ themselves. 
For a set $Y \subseteq A$, define
\[
\mathcal{B}(x;Y) := \bigcup_{y \in Y} \mathcal{B}(x,y).
\]
\end{definition}

\noindent
Intuitively, $\mathcal{B}(x;Y)$ collects all alternatives that lie along the metric line segments connecting $x$ to each element of~$Y$.
}


We can now extend Def.~\ref{def:safe_live_partial} to measure safety in any domain,  with the appropriate between set $\calB$. The difference from Def.~\ref{def:safe_live_partial} is colored in {\color{darkgreen}dark green}.


\begin{definition}[Safety, general domain]\label{def:safe_general}

 $\calR$ is safe with respect to $\calG$ and $V=S\cup H$  if $$\calR(V^+)\in {\color{darkgreen}\calB(}r;\calG(H){\color{darkgreen})}\ .$$
\end{definition}

We also extend the definition of liveness, by allowing honest voters to vote arbitrarily in $H_{\rightarrow}$. This is similar to the difference between a voting rule being \emph{unanimous} and being \emph{onto} (difference from Def.~\ref{def:safe_live_partial} is highlighted):
\begin{definition}[Liveness, general domain]\label{def:live_general}
  A voting rule $\calR$ is \emph{live} w.r.t  population $V=S\cup H$, if for all $a\in A$,  {\color{darkgreen}there is some alternative vote $H_{\rightarrow}$} of the honest voters such that $\calR(S\cup H^+_{\rightarrow}) = a$.
\end{definition}

The definitions above allow us to analyze different social choice settings; in the next subsections, we consider the following social choice settings -- below we mention what the between set means for each of them:
\begin{itemize}

\item
Multiple alternatives: That is, a discrete unordered set $A$. Here $\calB(x,y)=\{x,y\}$ as in the binary setting;
      
\item
Multiple referenda: with $d$ binary issues and the Hamming distance, i.e. $A=\{0,1\}^d$. Then $\calB(x,y)$ is the smallest \emph{box} containing both $x$ and $y$~\cite{nehring2007structure};
       
\item
Single-peaked preferences on lines: $\calB(x,y)$  is the smallest \emph{interval} containing both $x$ and $y$;

\item
Single-peaked preferences on trees: $\calB(x,y)$  contains all nodes in  the unique \emph{path} from  $x$ to $y$;
\end{itemize}  

\begin{remark}
Note that in the first case with two or more unordered alternatives, the general definitions of safety and liveness collapse to the simple ones we used in the previous sections (Def.~\ref{def:safe_live_partial}).
\end{remark}

\subsection{Multiple Alternatives}\label{section:beyond ma}

Here we consider setting in which $A$ is the set of alternatives with $r \in A$ being the status quo, but in which $|A| > 2$.
In contrast to the binary domain, where the Majority rule is the natural base rule, when $|A|>2$ there are many reasonable voting rules in the literature.  We start by extending some of our results to Plurality voting, then considering other voting rules.  

\paragraph{Plurality}
We can naturally extend the $\tSQMJ$ mechanism,  by using the Plurality rule $\calR=\PL$. That is, the mechanism $\tREPL$ applies the Plurality rule after adding a fraction of $\tau$ voters to $r$.

\begin{figure}
    \centering
\begin{tikzpicture}[scale=0.4]

\tikzstyle{dot}=[rectangle,draw=black,fill=white,inner sep=0pt,minimum size=4mm]
\tikzstyle{active}=[circle,draw=blue,text=white,fill=blue,inner sep=0pt,minimum size=3mm]
\tikzstyle{inactive}=[circle,draw=blue,fill=white,inner sep=0pt,minimum size=3mm]
\tikzstyle{sybil}=[rectangle,draw=black,fill=red,text=white,inner sep=0pt,minimum size=3mm]
\tikzstyle{del}=[triangle,draw=blue,fill=white,inner sep=0pt,minimum size=2mm]
\tikzstyle{virtual}=[diamond,draw=black,fill=gray,inner sep=0pt,minimum size=3mm]
\tikzstyle{txt}=[text width = 2cm, anchor=west]

\node at (3.5,4) {\textit{safety violation}};

\draw (-1,0.4) -- (10,0.4);
\node at (1,-.2) {$r$};
\node at (5,-.2) {$p$};
\node at (9,-.2) {$p'$};

\node at (5-0.5,2) [active] {};
\node at (5+0.5,2) [active] {};

\node at (9,2) [active] {};

\node at (1,2) {$\tau<0.6$};

\node at (1-0.5,1) [virtual] {};
\node at (1+0.5,1) [virtual] {};

\node at (9-0.5,3) [sybil] {};
\node at (9+0.5,3) [sybil] {};

\end{tikzpicture}
~~~~~~~~~~~
\begin{tikzpicture}[scale=0.4]

\tikzstyle{dot}=[rectangle,draw=black,fill=white,inner sep=0pt,minimum size=4mm]
\tikzstyle{active}=[circle,draw=blue,text=white,fill=blue,inner sep=0pt,minimum size=3mm]
\tikzstyle{inactive}=[circle,draw=blue,fill=white,inner sep=0pt,minimum size=3mm]
\tikzstyle{sybil}=[rectangle,draw=black,fill=red,text=white,inner sep=0pt,minimum size=3mm]
\tikzstyle{del}=[triangle,draw=blue,fill=white,inner sep=0pt,minimum size=2mm]
\tikzstyle{virtual}=[diamond,draw=black,fill=gray,inner sep=0pt,minimum size=3mm]
\tikzstyle{txt}=[text width = 2cm, anchor=west]

\node at (3.5,4) {\textit{liveness violation}};

\draw (-1,0.4) -- (10,0.4);
\node at (1,-.2) {$r$};
\node at (5,-.2) {$p$};
\node at (9,-.2) {$p'$};

\node at (5-0.5,2) [active] {};
\node at (5+0.5,2) [active] {};

\node at (9,2) [active] {};

\node at (1,2) {$\tau\geq0.6$};

\node at (1-1,1) [virtual] {};
\node at (1,1) [virtual] {};
\node at (1+1,1) [virtual] {};

\node at (1-0.5,3) [sybil] {};
\node at (1+0.5,3) [sybil] {};

\end{tikzpicture}
    \caption{We consider two instances with five active votes. On the left there is an instance where any $\tREPL$ mechanism with less than $3=0.6\cdot|V^+|$ virtual votes violates safety, since $p'$ is selected. On the right there is another instance where at least $3$ virtual voters mean violation of liveness since $r$ is selected regardless of how honest voters vote. }
    \label{fig:plu}
\end{figure}
\begin{observation} $\tREPL$ cannot be both safe  with respect to Plurality and live for three alternatives. This is regardless of $\tau$, and even if there is full participation ($\mu=0$) and only $\sigma>0.2$ sybils. 
\end{observation}

To see why, let $\eps\in (0,(\sigma-0.2))/2)$. Consider candidates $\{r,p,p'\}$ and suppose that $h_p=0.4$ honest voters vote $p$, and all other voters vote $p'$.  Thus $h_{p'}=1-h_p-\sigma<1-0.4-0.2=h_p$ and $p$ is the truthful outcome. A safe rule must therefore select $\tREPL(V)\in \calB(r;p) = \{r,p\}$.

Since $v_{p'}=h_{p'}+\sigma=0.6>v_p$, we get that  $p'$  is selected (which violates safety, see Fig.~\ref{fig:plu}, Left), unless $\tau\geq 0.6$. However if $\tau\geq 0.6$ then in a profile where all $\sigma$ vote $r$ there are $\tau+\sigma > 0.6+0.2 = 0.8 > h$ so neither $p$ nor $p'$ can be selected, regardless of how honest voters vote---i.e. liveness is violated. 

\medskip
The bound of $0.2$ is not tight, but instead of trying to characterize exactly the (deteriorated) safety-liveness tradeoff of $\tREPL$, we 
return to the $\tSM$ rule (see Def.~\ref{def:SMJ}). Its natural extension to multiple alternative is to select the unique alternative with strictly more than $\frac12+\tau$ votes, if one exists, and otherwise return $r$.

It turns out that when there are more than 2 alternatives, the  mechanism \emph{no longer coincides} with $\tREPL$. Moreover, $\tSM$ inherits the same safety and liveness guarantees from the binary case, whereas the example above shows that $\tREPL$ does not. 

\begin{theorem}
\label{thm:SMJ_safe}
  The $\tSM$ voting rule is safe w.r.t Plurality  if and only if $\tau\geq \frac{\sigma+\mu}{1-\mu}$.
\end{theorem}
\begin{theorem}\label{thm:SMJ_live}
   The $\tSM$ voting rule is live if and only if $\tau< \frac{1-2\sigma-\mu}{1-\mu}$.
 \end{theorem}
Note that the bounds in the theorems are identical to the bounds for $\tSM$ in the binary case (Section~\ref{sec:partial}), 
 which are the same bounds as  $\tSQMJ$. 
Theorem~\ref{thm:SMJ_safe} follows as a special case from Theorem~\ref{thm:SMJ_safe_alpha} in Section~\ref{sec:relaxed_domains}. 
For liveness, the number of alternatives is irrelevant so the proof of the binary case immediately applies for Theorem~\ref{thm:SMJ_live}.

In particular, obtaining both safety and liveness is possible iff $3\sigma+2\mu< 1$ (i.e. just as in the binary case).


Note that in the example above where $\tREPL$ fails (with $0.4$ of voters on $p$ and the rest on $p'$), using e.g. $0.3\shyp\SM$ is safe, since $v_{p'}=0.6 < 0.65 = (1+\tau)/2$, and thus $0.3\shyp\SM(V)=r$.

Another feature of  the $\tSM$ rule is that it may select $r$ even if no one voted for it!

\paragraph{Condorcet Conservative rules} Both Plurality and Supermajority allow only a simple ballot where every voter votes for a single alternative (plurality/$1$-approval ballots). 

However there are many other rules that are based on ranking the alternatives (i.e., voting rules for ordinal-based elections), such as Borda and other positional scoring rules, Maximin, STV and so on. 
Many voting rules are guided or justified by selecting the Condorcet winner, when one exists. The outcome of these rules typically differ when there is no Condorcet winner. 

A  `conservative' decision in the current context,  would mean selecting the status quo $r$ whenever there is no Condorcet winner. We call this rule the  \emph{Condorcet Conservative} rule ($\CC$).

The $\tau$-Super Condorcet Conservative rule ($\tau\shyp\SCC$) is similar but $p_i$ only beats $p_j$ if it has a supermajority of $\frac{1+\tau}{2}$ of the votes. That is, if there is an alternative $p$ that has a supermajority against any other alternative (including $r$) it is selected, and otherwise $r$ is selected. 

\begin{proposition}\label{prop:SCC}
The following hold:
\begin{itemize}
    \item  $\tau$-$\SCC$ has the same liveness guarantees as $\tSM$.
    \item  Let $\calG$ be any Condorcet consistent rule. Then $\tau$-$\SCC$ has the same safety guarantees with respect to $\calG$,  as $\tSM$ has with respect to $\MJ$.
\end{itemize}
 \end{proposition}
 \begin{proof}
 We prove each claim separately.
 
 \smallskip\noindent\textbf{Liveness:}\quad
 %
 
 Let $\tau,\mu,\sigma\geq 0$ such that $\tSM$ is live, and consider some $p\in A$. Set $H_{\rightarrow}$ s.t. all voters rank $p$ at the top.  In particular, when comparing $p$ to any other alternative $p'$ (including $r$), all honest voters vote for $p$ and thus liveness of $\tSM$ entails that $p$ is selected, i.e. has the required $\tau$-supermajority over $p'$. Thus $\tau\shyp\SCC(S\cup H^+_{\rightarrow}) = p$.

 \smallskip\noindent\textbf{Safety:}\quad 
 
 Let $\tau,\mu,\sigma\geq 0$ such that $\tSM$ is safe.
 Consider any profile $V=H\cup S$ where some $p\neq r$ wins in $\tau\shyp\SCC(V^+)$ (otherwise safety is trivial). Then we need to show that $\calG(H)=p$.
 
 Indeed, consider any $p'\neq p$ (including $r$). 
 Since $\tau\shyp\SCC(V^+)=p$, we know that in the pairwise match of $p$ vs. $p'$, there is a fraction of at least $(\frac{1+\tau}2)v^+$ voters that prefer $p$, meaning that $p$ beats $p'$ under $\tSM$. 
 
 By safety of $\tSM$ (and since $p\neq r$), this means that more than half of the honest voters prefer $p$ over $p'$. Since this holds for all $p'\neq p$, we have that $p$ is the Condorcet winner of $H$, and thus $\calG(H)=p$.
 \end{proof}
 
An immediate implication of Prop.~\ref{prop:SCC} is that the bounds of Theorem \ref{thm:SQMJ} 
 hold also for the $\tau\shyp\SCC$ rule.

\subsection{Multiple Referenda}\label{section:beyond mr}

We move to the social choice setting of multiple referenda. That is, suppose that $A=\{0,1\}^d$, where w.l.o.g. $r=\boldsymbol{0}$. For a base rule, we use the issue-wise Majority rule $\IMJ$, which simply selects the majority opinion on each of the $d$ issues (this is a \emph{combinatorial domain}~\cite{combinatorialdomains}). Note that $\IMJ(U)$ minimizes the sum of Hamming distances to all voters in $U$, thus maximizing the standard definition of the social welfare. 

\begin{proposition}\label{prop:IMJ}
The following hold:
\begin{itemize}
    \item  $\tau\shyp\IMJ$ has the same liveness guarantees as $\tSQMJ$.
    \item   $\tau\shyp\IMJ$ has the same safety guarantees with respect to $\IMJ$,  as $\tSQMJ$ has with respect to $\MJ$.
\end{itemize}
 \end{proposition}
 \begin{proof}
  For an issue $j\leq d$ and voter set $U$, we denote by $U|_j\in\{0,1\}^{|U|}$ the projected opinions of all $U$ voters on issue $j$.
We prove each claim separately.

 \smallskip\noindent\textbf{Liveness:}\quad
 Let $\tau,\mu,\sigma\geq 0$ such that $\tSQMJ$ is live.
Consider some position $p\in \{0,1\}^d$. For any given profile $V=H\cup S$, set $H_\rightarrow$ s.t. all honest voters vote for $p$. This means that in $S\cup H_\rightarrow$ at least $h^+$ honest voters agree with  $p_j$ for every issue $j$.
From liveness it follows that $\tSQMJ((S\cup H^+_\rightarrow)|_j)=p_j$.
Thus 
$\tREIMJ(S\cup H^+_\rightarrow)=(p_j)_{j\leq d}= p$.

 \smallskip\noindent\textbf{Safety:}\quad Let $\tau,\mu,\sigma\geq 0$ such that $\tSQMJ$ is safe w.r.t $\MJ$.
 Suppose that $\tREIMJ(V^+)=p \neq r$ (otherwise $0$-safety is trivial). To show safety, we need to prove $p\in \calB(r;\IMJ(H))$.\footnote{This is the first nontrivial use of the ``betweenness" notion in the paper, i.e. where the set contains not just $r$ and $\calG(H)$.  See Definition~\ref{definition:between}.}
 This means showing $p_j \in \{r_j,\IMJ(H)_j\}$ for all $j\leq d$.
 
 By safety of $\tSQMJ$, we know that $\tSQMJ(V^+|_j)\in \{r_j,\MJ(H|_j)\}$ for all $j$. To complete the proof, we observe that $p_j=\tREIMJ(V^+)_j = \tSQMJ(V^+|_j)$ and that $\{r_j,\IMJ(H)_j\}=\{r_j,\MJ(H|_j)\}$. 
\end{proof}

As with the Condorcet Conservative rule, we can conclude that the bounds in Theorem~\ref{thm:SQMJ} 
apply to $\tREIMJ$.

\subsect{Single-peaked Domains}\label{section:oned}
%
\revision{
In this section we consider voters that have single peaked preferences on a line (discrete or continuous) or a tree. 

Single-peaked preferences on a line were first considered by Black~\cite{black1948rationale}. For extensions to trees and other domains see \cite{nehring2007structure}. We introduce here an equivalent definition on the notion of betweenness (see Def.~\ref{definition:between}).

\begin{definition}[Single-peaked domains]
 A metric space $(A,\delta)$ is single-peaked if for any two positions $x,y\in A$, a voter at $x$  prefers any $z\in \calB(x,y)$ over $y$.  
\end{definition}

The single-peaked domains we will consider in this section are lines and trees. It is well known that in these domains, any profile of voters has a \emph{median} (unique if the number of voters is odd): a point from which moving to either side takes us farther away from most voters~\cite{nehring2007structure}. 

The \emph{Median voting rule} ($\MD$) returns this point, using some lexicographic tie-breaking rule if needed. The median rule has many desired properties such as Condorcet consistency, strategyproofness, and social optimality~\cite{black1948rationale,moulin1980strategy,procaccia2009approximate, nehring2007structure}, and it therefore makes sense to apply it as our base rule $\calG$.\footnote{In the Appendix, in Section~\ref{section:mean} we also consider the mean on lines, generally showing its  (rather expected) inferiority to the median.}

%

Cohensius et al.~\shortcite{cohensius2017proxy} consider the case of a line with a small fraction of active participants but with no sybils,\footnote{The opposite case of sybils with full participation was considered in the two conference papers initiating the current work: in \cite{srsc} a specialized voting rule that explicitly ignores the most extreme votes was suggested and analyzed; and in \cite{srscpp} we showed that this rule in fact coincides with $\tSQMD$. We therefore only consider $\tSQMD$ here.}  and we return to their model in Section~\ref{sec:delegation}.

As in the previous sections, we consider the $\tSQMD$ rule which places $\tau|V^+|$ virtual voters on the status quo $r$, and analyze its safety and liveness guarantees using a reduction to the binary setting. 


We consider an arbitrary population $V=H^+ \cup H^- \cup S$ with partial participation and sybils and consider $\tSQMD$.
\full{
We use the following straightforward connection between the median and majority rules.
\begin{lemma}\label{lemma:MD_MJ}
  Let $z$ be the position of the median voter of $V$, and let $x\neq y$ s.t. $y$ is between $x$ and $z$. Then $y$ has a  majority in $V$ against $x$.
\end{lemma}
This is simply because for every voter $i$ s.t. $z$ is between $x$ and $a_i$ (at least half the voters),  $y$ is also between $x$ and $a_i$. See Fig.~\ref{fig:tree_median}.

The lemma clearly still holds if we modify the set of voters by adding votes for $r$ and/or ignoring passive voters. 
Thus, the lemma still applies if we replace ``median'' with $\tSQMD$  and ``majority'' with $\tSQMJ$,  or replace $V$ with $V^+$.
We use Lemma~\ref{lemma:MD_MJ} to derive the following.
}
\begin{figure}
    \centering

\begin{tikzpicture}[
    scale=1.5,
    vertex/.style={circle, draw, minimum size=0.6cm, inner sep=0pt, fill=white},
    agent/.style={circle, draw, fill=lightgray, minimum size=0.6cm, inner sep=0pt},
    special/.style={circle, draw, fill=yellow!30, minimum size=0.6cm, inner sep=0pt},
    edge/.style={thick}
]

\tikzstyle{dot}=[rectangle,draw=black,fill=white,inner sep=0pt,minimum size=4mm]
\tikzstyle{active}=[circle,draw=blue,fill=blue,inner sep=0pt,minimum size=2mm]
\tikzstyle{inactive}=[circle,draw=blue,fill=white,inner sep=0pt,minimum size=2mm]
\tikzstyle{sybil}=[rectangle,draw=black,fill=red,inner sep=0pt,minimum size=1.6mm]
\tikzstyle{del}=[triangle,draw=blue,fill=white,inner sep=0pt,minimum size=2mm]
\tikzstyle{virtual}=[diamond,draw=black,fill=gray,inner sep=0pt,minimum size=2mm]
\tikzstyle{txt}=[text width = 8cm, anchor=west]

\node[special] (x) at (0, 0) {$y$};  
\node[special] (v1) at (-2, 1) {$x$};
\node[vertex] (v2) at (0, 1) { };
\node[vertex] (y) at (2, 1) { };  
\node[vertex] (v4) at (-3, 2) { };
\node[active]  at (-3,2) { }; 
\node[vertex] (a1) at (-2, 2) { };  
\node[active]  at (-2.1,2) { }; 
\node[active]  at (-1.9,2) { }; 
\node[vertex] (a2) at (-1, 2) { };  
\node[vertex] (a3) at (0, 2) { };  
\node[active]  at (0,2) { }; 
\node[vertex] (v8) at (0,3) { };
\node[special] (z) at (2, 2) {$z$};  
\node[vertex] (v9) at (-4, 3) { };
\node[vertex] (a4) at (-3, 3) { };  
\node[active]  at (-3,3) { }; 
\node[vertex] (a5) at (2, 3) { };  
\node[active]  at (2,3) { }; 

\draw[edge] (x) -- (v1);
\draw[edge] (x) -- (v2);
\draw[edge] (x) -- (y);
\draw[edge] (v1) -- (v4);
\draw[edge] (v1) -- (a1);
\draw[edge] (v1) -- (a2);
\draw[edge] (v2) -- (a3);
\draw[edge] (v8) -- (a3);
\draw[edge] (y) -- (z);
\draw[edge] (v4) -- (v9);
\draw[edge] (v4) -- (a4);
\draw[edge] (z) -- (a5);

\node[vertex] at (-1,0.5) { };
\node[vertex] at (1,0.5) { };
\node[active]  at (1,0.5) { }; 

\end{tikzpicture}
\caption{An example of seven voters on a tree. $x$ is the tree median, so by definition it is preferred by a majority of voters to any other point. In particular, the four voters to the top-left prefer $x$ over $z$, and therefore must also prefer $y$ over $z$. }
    \label{fig:tree_median}
\end{figure}
\begin{theorem}\label{thm:MJ_MD_reduction}
The following hold:
\begin{itemize}
    \item  $\tSQMD$ has the same liveness guarantees as $\tSQMJ$.
    \item   $\tSQMD$ has the same safety guarantees with respect to $\MD$,  as $\tSQMJ$ has with respect to $\MJ$.
\end{itemize}
\end{theorem}
Clearly, if $\tSQMJ$ violates safety/liveness in some profile $V$, create an instance where all voters are located either on $r$ or on $p:=\tSQMD(V)$ (according to their preference in $V$). Then, $\tSQMD(V)=\tSQMJ(V)$ so we get a violation of safety/liveness in $\tSQMD$ as well.

In the other direction, the construction is somewhat more involved.
The proof for safety will follow from the more general Theorem~\ref{thm:MJ_MD_reduction_alpha}, which also considers approximate safety.
\begin{proof}[Proof for liveness]
 For a profile $U$ of locations on $\mathbb R$ and a pair of locations $x,y\in \mathbb R$, we denote by $U|_{xy}$ the projection of $U$ on $A=\{x,y\}$. That is, a binary profile where each voter votes for the more preferred alternative among $x$  and $y$. In case of a tie, the voter selects $x$.

\medskip
 Consider any set of parameters $\mu,\sigma,\tau\geq 0$ such that $\tSQMD$ is live. Let $V=S\cup H$ be some profile of voters, and let $x$ be any position on the line or tree. We argue that $\tSQMD(S\cup H^+_{\rightarrow x})=x$.

 Indeed, denote $V^x=S\cup H^+_{\rightarrow x}$ and assume towards a contradiction that $\tSQMD(V^x)=y\neq x$. 

Note that in the binary profile $V^x|_{xy}$ all honest voters vote for $x$, thus from liveness of $\tSQMJ$ we get $\tSQMJ(V^x|_{xy})=x$. On the other hand, since the median of $V^x$ is at $\tSQMD(V^x)=y \neq x$, then by Lemma~\ref{lemma:MD_MJ}, $y$ has a majority against $x$, which is a contradiction. 

The other direction is trivial, by considering profiles where all voters are restricted to $r$ and one other position.
\end{proof}

The above reduction allows us to easily transfer all previous results to single-peaked domains. 
\begin{corollary}\label{cor:MD_tradeoff}
The following hold under single-peaked preferences on lines and trees:
\begin{itemize}

\item
$\tSQMD$ is safe w.r.t $\MD$ as the base rule if and only if $\tau\geq \frac{\sigma+\mu}{1-\mu}$.
\item $\tSQMD$ is live iff $\tau< \frac{1-2\sigma-\mu}{1-\mu}$.
\item 
There is no mechanism $\calR$ that is both safe w.r.t $\MD$ and live when $3\sigma+2\mu\geq 1$.
\end{itemize}
\end{corollary}

}
\section{Relaxed Safety}\label{section:relaxed}
Consider the simplest setting with Majority as our base rule.
So far we have treated safety  as a dichotomy: for a given fraction of sybils, a mechanism is either safe or not. 

However, if we think about violation of safety  as a situation in  which most honest voters prefer the status quo $r$  and the mechanism (perhaps due to sybils or abstention) selects  $p$, then it should also be clear that some violations are worse than others:
\begin{itemize}
    \item If the honest voters are almost evenly split between $r$ and $p$ then it does not matter much which alternative is selected, as both outcomes are `acceptable';
    \item  In contrast, if there is an overwhelming majority of honest voters for $r$ (meaning only $r$ is acceptable) but $p$ is selected then this is a more serious violation of safety.
\end{itemize}   

Next, we introduce a formal definition of an acceptable outcome that contains a sensitivity parameter. 

\begin{figure*}[t]
    \centering
 \begin{tikzpicture}[scale=0.8]

\tikzstyle{dot}=[rectangle,draw=black,fill=white,inner sep=0pt,minimum size=4mm]
\tikzstyle{active}=[circle,draw=blue,fill=blue,inner sep=0pt,minimum size=2mm]
\tikzstyle{inactive}=[circle,draw=blue,fill=white,inner sep=0pt,minimum size=2mm]
\tikzstyle{sybil}=[rectangle,draw=black,fill=red,inner sep=0pt,minimum size=1.6mm]
\tikzstyle{del}=[triangle,draw=blue,fill=white,inner sep=0pt,minimum size=2mm]
\tikzstyle{virtual}=[diamond,draw=black,fill=gray,inner sep=0pt,minimum size=2mm]
\tikzstyle{txt}=[text width = 8cm, anchor=west]

\node at (0,-0.5) {$r=0$};
\node at (0,0) {$*$};
\node at (5,0) {$*$};
\node at (5,-0.5) {$\calG(H)=3$};

\draw[ultra thin] (6.5,0) -- (-0.8,0);
\draw[thick,color=darkgreen] (3.6,0.05) -- (5.8,0.05);
\draw[dashed, thick,color=darkgreen] (3.6,0.05) -- (0,0.05);

\draw [decorate,decoration={brace,amplitude=4pt,raise=4pt},yshift=0pt]
(3.6,0.1) -- (5.8,0.1) node [black,midway,yshift=0.8cm] {
$\overline \calG_\safe(H)$};
\end{tikzpicture}
~~~~~~~~~
 \begin{tikzpicture}[scale=0.85]

\tikzstyle{dot}=[rectangle,draw=black,fill=white,inner sep=0pt,minimum size=4mm]
\tikzstyle{active}=[circle,draw=blue,fill=blue,inner sep=0pt,minimum size=2mm]
\tikzstyle{inactive}=[circle,draw=blue,fill=white,inner sep=0pt,minimum size=2mm]
\tikzstyle{sybil}=[rectangle,draw=black,fill=red,inner sep=0pt,minimum size=1.6mm]
\tikzstyle{del}=[triangle,draw=blue,fill=white,inner sep=0pt,minimum size=2mm]
\tikzstyle{virtual}=[diamond,draw=black,fill=gray,inner sep=0pt,minimum size=2mm]
\tikzstyle{txt}=[text width = 8cm, anchor=west]

\node at (-0.3,0) {$r$};
\node at (0,0) {$*$};
\node at (5,2) {$*$};
\node at (5.65,2) {$\calG(H)$};

\draw [decorate,decoration={brace,amplitude=4pt,raise=4pt},yshift=0pt]
(6.5,1) -- (4.4,1.1-0.3666) node [black,midway,yshift=-0.6cm] {
$\overline \calG_\safe(H)$};

\draw (4,2.2) -- (4.8,2.9);
\draw[color=darkgreen] (4.8,2.9) -- (6.4,2.5);
\draw[color=darkgreen] (6.4,2.5) -- (5.8,1.1);
\draw[color=darkgreen] (5.8,1.1) -- (5.1,0.8);
\draw (5.1,0.8) -- (4.3,1.1);
\draw (4.3,1.1)-- (4,2.2) ;

\draw[dashed,color=darkgreen] (4.8,2.9) -- (0,0);
\draw[dashed,color=darkgreen]  (5.1,0.8) -- (0,0);
\end{tikzpicture}
    \caption{A demonstration of the $\safe$-safety property in the 1-dimensional real line (left) and in some 2-dimensional metric space (right). The status quo $r$ and the ideal point $\calG(H)$ are marked by $*$.  The area inside the solid line is $\ol \calG_\safe(H)$. The area inside in the green lines (solid or dashed) is $\calB(r;\ol\calG_\safe(H))$.
    }
    \label{fig:safety_1D2D}
\end{figure*}

\revision{
\paragraph{Outcome range and robustness to small perturbations}
To quantify how sensitive an aggregation rule is to small changes in participation or voting behavior, 
we define the notion of an \emph{outcome range}. 
This captures the set of possible outcomes that can result from altering the votes of only a limited fraction of the honest population.

\begin{definition}[Outcome range]\label{def:range}
Let $\mathcal{R}$ be an aggregation rule, and let the population be $V = H \cup S$. 
For a parameter $\gamma \ge 0$, define
\[
\overline{\mathcal{R}}_{\gamma}(V) := 
\left\{ 
\mathcal{R}(H' \cup S) \; : \; 
\exists H' \text{ with } |H'| \ge |H| \text{ and } |H' \setminus H| \le \gamma |H| 
\right\}.
\]
\end{definition}

\noindent
For $\gamma \in [0,1]$, the set $\overline{\mathcal{R}}_{\gamma}(V)$ contains all outcomes that can be obtained by 
replacing at most a $\gamma$-fraction of the honest voters with arbitrary votes.
In this sense, $\gamma$ measures the \emph{input robustness} of the rule.

\medskip
\noindent
When $\gamma = 0$, we recover the original outcome $\overline{\mathcal{R}}_{0}(V) = \mathcal{R}(V)$.
As $\gamma$ increases, the range enlarges, reflecting greater tolerance to perturbations.
This notion of approximation concerns the \emph{input side}—the fraction of voters that must change to alter the outcome—rather than the similarity between alternatives themselves.\footnote{
This perspective is sometimes called \emph{input approximation}, in contrast to \emph{output approximation}~\cite{meir2018strategic}. 
It can also be viewed as a negative analogue of the margin of victory: 
an alternative is considered acceptable if it could win after modifying only a small share of votes.
}
}

\medskip

\revision{
\paragraph{Outcome range under Majority}
In the binary setting, the outcome range depends only on how close the honest electorate is to a tie. 
Intuitively, if the honest votes are nearly balanced, then small perturbations—captured by the parameter~$\gamma$—may change the outcome, whereas if the margin is large, the outcome remains stable.

Suppose that $\MJ(H) = r$. 
Then $\overline{\MJ}_{\gamma}(H)$ necessarily includes $r$; the key question is when it also includes $p$.
}

\begin{observation} \label{ob:tie}
    In the binary setting, $p\in \ol \MJ_\range(H)$ if and only if $h_p > h_r -2\range\cdot h$; and $r\in \ol \MJ_\range(H)$ if and only if $h_r \geq h_p - 2\range \cdot h$.
\end{observation}

\begin{proof} We show this for $p$. The proof for $r$ is symmetric except for the tie-breaking. 

Suppose $h_p > h_r -2\range\cdot h$, then either $h_p > h_r$, in which case $p=\MJ(H) \in \ol\MJ_\range(H)$; or $h_r \geq 0.5h$. Set $\range':=\min\{0.5,\range\}$ then $\range'h\leq h_r$. Now, Let $H''\subseteq H_r$ be an arbitrary set of $r$ voters of size $\range'$, and let $H':=(H\setminus H'') \cup H''_{\rightarrow p}$. We then have 
\begin{align*}
    h'_p-h'_r &= h_p + \range'h - (h_r-\range'h) = h_p-h_r + 2\range'h \\
    & = \min\{ h_p-h_r + 2\range'h,  h_p-h_r + h\} \geq 0.
\end{align*}  
On the other hand, if $h_p \leq h_r -2\range\cdot h$, then $h_p+\range h \leq 0.5h \leq h_r-\range h$, and in any population $H'$ with a majority for $p$ we have 
\begin{align*}
|H'\setminus H| &\geq |H'_p\setminus H_p| = |H'_p|-|H_p| = |V|(h'_p-h_p)\geq |V|(h'_p - (h_r-2\range h))\\
&> |V|(0.5h' - (h_r-2\range h)) \geq |V|(0.5h - (h_r-2\range h)) \\
&= |V|(0.5h - (h_r-\range h)+ \range h) \geq |V|\range h = \range |H|,
\end{align*}
which means $p\notin \ol\MJ_\range(H)$.
\end{proof}

\paragraph{Quantifying Safety}
Following the above discussion, we extend the definition of safety with a parameter. We highlight the difference from Def.~\ref{def:safe_general} in {\color{red}red}.
\begin{definition}[Quantified safety]\label{def:safe_quant}
$\calR$ is {\color{red}{$\safe$}}-safe with respect to $\calG$ and $V=S\cup H$  if\\ $\calR(V^+)\in \calB(r;{\color{red}\ol\calG_\safe(H)})$.
\end{definition}
Note that for $\alpha=0$ the definition collapses to safety, as in Def.~\ref{def:safe_general}.

Fig.~\ref{fig:safety_1D2D} demonstrates how the outcome range combines with  the notion of betweenness in Euclidean spaces. The $\safe$-safe area $\calB(r;\ol \calG_\alpha(H))$ includes all alternatives enclosed in either dashed or solid lines.

\subsection{Relaxed Safety in the Binary Setting}\label{sec:relaxed_binary}

\revision{
So far, safety was treated as an all-or-nothing property: the mechanism was either safe or unsafe. 
In many applications, however, it is useful to quantify \emph{how} safe a rule is—that is, how far it can deviate from full safety while still maintaining bounded risk. 
We therefore introduce a relaxed version, parameterized by~$\alpha$, that measures the maximal deviation from the ideal safety condition.
}

Our next theorem characterises exactly the conditions in which $\tSQMJ$ is $\safe$-safe. This is also visualized (for specific values) in Fig.~\ref{fig:alpha-safe}.

\begin{figure}
    \begin{center}

 \begin{tikzpicture}
  \begin{axis}[
    xmin=0, xmax=1,
    ymin=0, ymax=2.3,
    axis lines=middle,
    xlabel={ abstention $\mu$},
        x label style={at={(axis description cs:0.5,-0.18)},anchor=south},
    ylabel={parameter $\tau$},
    y label style={at={(axis description cs:-0.1,.5)},rotate=90,anchor=south},
    y tick label style={
    /pgf/number format/.cd,
            fixed,
            fixed zerofill,
            precision=2,
        /tikz/.cd
        },
    xtick={0.12, 0.28,0.6,0.8},
    ytick={0.04,0.2,0.6,1,2},
    legend style={at={(axis cs:1,1.5)},anchor=north,legend cell align=left}
    ]
    
    \addplot[blue, thick, domain=0:0.65, samples=100, name path=f_safe] {(1+0.2)/(1-x)-1};
      \addlegendentry{safety}
    \path[name path=Haxis] (axis cs:0,2.1) -- (axis cs:0.61,2.1);
    \addplot[blue!30,semitransparent,forget plot] fill between[of=f_safe and Haxis, soft clip={domain=0:1}];
     \node[blue] at (axis cs: 0.25,1.2) {\LARGE{safe}};

 \addplot[blue, thin, domain=0:0.7, samples=100, name path=f_safe2] {(-0.1*2*(1-0.2)+(1+0.2))/(1-x)-1};
      \addlegendentry{$0.1$-safety}

       \addplot[blue, thin,dashed, domain=0:0.75, samples=100, name path=f_safe3] {(-0.2*2*(1-0.2)+(1+0.2))/(1-x)-1};
             \addlegendentry{$0.2$-safety}
                    \addplot[blue, thin,dotted, domain=0:0.75, samples=100, name path=f_safe4] {(-0.3*2*(1-0.2)+(1+0.2))/(1-x)-1};
             \addlegendentry{$0.3$-safety}

    \addplot[red, domain=0:0.99, samples=100, name path=f_live] {2*(1-0.2-x)/(1-x)-1};
                  \addlegendentry{liveness}
    \path[name path=Laxis] (axis cs:0,0) -- (axis cs:1,0);
    \addplot[red!30,semitransparent,forget plot] fill between[of=f_live and Laxis, soft clip={domain=0:1}];


  \end{axis}
\end{tikzpicture}
\end{center}
\caption{Visualization of relaxed safety on the same example from Fig.~\ref{fig:tradeoff}. Here the fraction of sybils is fixed at $\sigma=0.2$ and the different curves show the range of $\safe$-safe mechanisms for different levels of safety. Note that Majority is $\safe$-safe w.r.t. itself whenever the curve is below the X-axis. \revision{Parameters are as in Figure~\ref{fig:tradeoff}: $\tau$ controls the strength of the status quo bias, $\sigma$ the proportion of sybils, and $\mu$ the fraction of abstaining honest voters.}
    \label{fig:alpha-safe}}
\end{figure}

\begin{theorem}[Safety bound]\label{thm:SQMJ_safe_alpha}
  The $\tSQMJ$ voting rule is $\safe$-safe w.r.t Majority as the base rule if and only if $$\safe \geq\frac{1+\sigma-(1+\tau)(1-\mu)}{2(1-\sigma)}\ .$$
\end{theorem}

Note that the safety bound in Theorem~\ref{thm:SQMJ} is derived by setting $\alpha=0$.
\revision{
Before turning to the formal argument, recall that $\alpha$ quantifies how much deviation from perfect safety we are willing to tolerate. 
The bound below specifies the minimal level of such relaxation needed for the $\tau$-SQ-MJ mechanism to remain safe despite sybils and abstentions. 
As $\alpha$ increases, the permissible region of $(\sigma,\mu,\tau)$ values expands correspondingly, see also in Fig.~\ref{fig:alpha-safe}.
}

\begin{proof}
Consider a given profile $V$.
If $\tSQMJ(V^+)=r$ or $p\in \ol{\MJ}_\safe(H)$ then there is no violation of $\safe$-safety and we are done. Thus, assume that $\tSQMJ(V^+)=p\neq r$. Recall that $h^+_p$ denotes the fraction of active honest voters voting for $p$. W.l.o.g.\ we may assume that all of $S$ vote for~$p$, since if profile $V$ violates $\safe$-safety, we can define a new profile $V'$, by switching  all $S$ agents who vote for $r$ with $p$ voters, and we would still have $\tSQMJ(V^+)=p$ (and $\ol{\MJ}_\safe(H) $  is unaffected) and thus there is still a violation in $V'$ (so, intuitively, profiles in which all sybils vote for $p$ are the hardest case for keeping safety). Similarly, we assume w.l.o.g that all of $H^-$ vote for $r$, thus $h_r=h^-+h^+_r,\ h_p=h^+_p$ (again,  profiles in which all passive voters vote for $r$ are the hardest case for keeping safety, as safety is defined w.r.t all honest voters); so, the fraction of active honest voters voting for $r$ is $h^+_r=1-\sigma-\mu-h^+_p$. Since $\tSQMJ(V^+)=p$, we have that
\begin{align}
 h^+_p+\sigma & = v^+_p > v^+_r+q = h^+_r + q = h^+-h^+_p +q \notag\\ &=(1-\mu-\sigma-h^+_p)+\tau(1-\mu)\ \text{, and thus}\notag\\
 2h^+_p &> (1+\tau)(1-\mu)-2\sigma. \label{eq:x_bound_RE} 
\end{align}


To show that $p\in \ol{\MJ}_\safe(H)$, which would show $\safe$-safety, it is left to show that we can change the votes of $\safe \cdot |H|$
honest voters from $r$ to $p$, to create a new profile $H'$ where $p$ has a strict majority of honest votes. 
Denote
 \begin{equation}\label{eq:beta_tag}
 {\safe'=\safe h = \safe(1-\sigma)\geq \frac{1+\sigma-(1+\tau)(1-\mu)}{2}\ .}
 \end{equation}
Indeed, after moving $\safe'$ votes, $r$ has 
$$h'_r = h_r-\safe' = h-h_p-\safe' = 1-\sigma-h^+_p -\safe'$$ 
honest votes, whereas $p$ has $h'_p=h^+_p+\safe'$ honest votes.
Therefore,  we have that
\begin{align*}
h'_p-h'_r & = (h^+_p+\safe')-(1-\sigma-h^+_p-\safe') \\
&= 2(h^+_p +\safe') - (1-\sigma) \\
&\geq 2h^+_p +  (1+\sigma-(1+\tau)(1-\mu)) - (1-\sigma)\tag{By Eq.~\eqref{eq:beta_tag}}\\
&> (1-\sigma)-(1-\sigma) = 0\tag{By Eq.~\eqref{eq:x_bound_RE}} .
\end{align*}
So, there are strictly more honest votes for $p$ than for $r$.

\medskip
In the other direction (i.e. to show tightness of the bound), consider $\tau,\sigma,\mu$ and $\safe<\frac{1+\sigma-(1+\tau)(1-\mu)}{2(1-\sigma)}$:
First set $\eps=\frac{1+\sigma-(1+\tau)(1-\mu)}{2(1-\sigma)}-\safe$.
Next, set $h^+_p = \frac{(1+\tau)(1-\mu)-2\sigma}{2}+\eps'$, where $\eps'\in(0,\frac{\eps}{1-\sigma})$. All $\sigma$ sybils vote for $p$, and all $\mu$ inactive honest voters vote for $r$. 

It is left to show that (a) $\ol{\MJ}_\safe(H)=\{r\}$ \full{(i.e. $r$ is the only safe outcome)}; and that (b) $\tSQMJ(V^+)=p$ (details omitted).
\full{For (a), consider any honest profile $H'$ such that $|H'\setminus H|\leq \safe |H|$. In the best case, we have that $h'_p \leq h_p+\safe h$ and $h'_r \geq h_r-\safe h$.
Indeed,
\begin{align*}
h'_p&-h'_r \leq h_p-h_r+2\safe h = h_p-(h-h_p)+ 2 \safe h\\
&= 2h_p-(1-\sigma) + 2\safe(1-\sigma)\\
&= 2h^+_p-(1-\sigma) + 2\safe(1-\sigma)  \\ 
& = [(1+\tau)(1-\mu)-2\sigma+2\eps'] - (1-\sigma) \\
&~~~+ [(1+\sigma)-(1+\tau)(1-\mu) + 2\eps(1-\sigma)]\\
& = 2\eps' -2\eps(1-\sigma) <0,
\end{align*}
which shows that $MJ(H')=r$ as required.

For (b), we can see that 
\begin{align*}
    v^+_p &- (v^+_r+q)  = (h^+_p+\sigma) - ((h^+-h^+_p) +  \tau v^+) \\
    & = 2h^+_p  - h^+ - v^+\tau + \sigma\\
    &= 2h^+_p - (1-\sigma-\mu) - (1-\mu)\tau - \sigma\\
    & = 2h^+_p - (1-\mu)(1+\tau) -2\sigma \\
   &= 2\eps' > 0, \tag{by definition of $h^+_p$}
\end{align*}
which shows that $\tSQMJ(V^+)=p$ and thereby completes the proof.}
\end{proof}

\paragraph{Mechanism design perspective}
The analysis of the $\safe$-safety of $\tSQMJ$ for given values of $\sigma$ and $\mu$ implies a different point of view:
  Indeed, in practical situations, the value of $\safe$-safety might be decided by a user of the system (a stricter user would require smaller values);
  then, given some estimations of $\sigma$ and $\mu$ ($\mu$ is usually known exactly since we know who is eligible to vote, while to estimate $\sigma$ one can use, e.g., sampling techniques can be used to infer what value of $\tau$), the user shall choose for the $\tSQMJ$ mechanism to achieve the desired level of safety. 
  
  For example, we can see in  Fig.~\ref{fig:alpha-safe} that under $\sigma=0.2$ and $\mu=0.3$ it would not be possible to get both liveness and full safety, but $0.2$-safety can still be obtained. Refer to Section~\ref{section:discussion} for a further discussion on the mechanism design perspective.

\subsection{Relaxed Safety in Other Domains}\label{sec:relaxed_domains}

Some of the safety bounds  for the domains studied in Section~\ref{sec:domains} similarly generalize to any $\alpha\geq 0$, as they are essentially based on a reduction to the binary domain that preserves the approximation. These include the results for multiple alternatives and single-peaked domains. In contrast, our results from multiple referenda and Condorcet-conservative rules do not generalize to arbitrary $\alpha$. 

\paragraph{Multiple alternatives}
\begin{theorem}
\label{thm:SMJ_safe_alpha}
  The $\tSM$ voting rule is $\safe$-safe w.r.t Plurality  if and only if $\safe \geq\frac{1+\sigma-(1+\tau)(1-\mu)}{2(1-\sigma)}$.
\end{theorem}
By setting $\alpha=0$ we get Theorem~\ref{thm:SMJ_safe}.

\begin{proof}
We follow the same steps as in the proof of Theorem~\ref{thm:SQMJ_safe_alpha}: 
Suppose that $\tSM$ selects $p$, then we need to show $p$ is $\safe$-safe by making it the honest winner. That is, we need to construct a modified profile $H'$ where $p$ has most votes.  In fact, we will show it gets a strict majority.
For this, we need to provide corresponding inequalities to Eqs.~\eqref{eq:x_bound_RE} and \eqref{eq:beta_tag}.

For the first, we observe that in $\tSM(V^+)$, alternative $p$ gets more than $(1+\tau)/2$ of all active votes.\footnote{This is exactly where the proof would fail for $\tREPL^+$, since $p$ can win even with a lower fraction of votes.} Thus 
\begin{align}
    &h^+_p+\sigma \geq v^+_p > \frac{1+\tau}{2}v^+ = \frac{1+\tau}{2}(1-\mu) \Rightarrow \notag\\
    & 2h^+_p > (1+\tau)(1-\mu)-2\sigma \label{eq:x_bound_RE_SMJ}.
\end{align}

Now, set 
\labeq{beta_tag_SMJ}
{\safe'=\safe h \geq \frac{1+\sigma-(1+\tau)(1-\mu)}{2}\ .}

Then, to construct $H'$, we move a fraction of $\safe$ honest voters to $p$, from any other alternative (not necessarily from $r$). We get: 

 \begin{align*}
2h'_p-h &= 2(h^+_p +\safe') - (1-\sigma) \\
&\geq 2h^+_p +  (1+\sigma-(1+\tau)(1-\mu)) - (1-\sigma)\tag{By Eq.~\eqref{eq:beta_tag_SMJ}}\\
&> (1-\sigma)-(1-\sigma) = 0\tag{By Eq.~\eqref{eq:x_bound_RE_SMJ}}\ ,
\end{align*}
so $h'_p> 0.5h$, as required.
\end{proof}

\revision{
\paragraph{Lines and trees}

Here, we concentrate on the  median rule; in Appendix~\ref{section:mean} we consider the mean (on a line) as well, generally showing its (rather expected) inferiority.

\begin{theorem}\label{thm:MJ_MD_reduction_alpha}
$\tSQMD$ has the same safety guarantees with respect to $\MD$,  as $\tSQMJ$ has with respect to $\MJ$, for any $\alpha\geq 0$.   
\end{theorem}
\rmr{I wrote a new proof and added a graphical example on a tree}
\begin{proof}
    We first show that $\safe$-safety on a tree entails $\safe$-safety in the binary setting.  Indeed, assume that for some set of parameters $\mu,\sigma,\tau,\safe\geq 0$, $\tSQMJ$ is \emph{not} $\safe$-safe w.r.t. simple majority. In particular this means there is a profile $V=H^+\cup H^-\cup S$ on $\{r,p\}$ s.t. $\ol \MJ_\alpha(H) = \{r\}$ but $\tSQMJ(V^+) = \MJ(S\cup H^+\cup Q)=p$. 
    
    Pick two arbitrary points on the tree or line, label them $r$ and $p$, and place each voter from the binary instance above in its respective position, getting a new profile $\hat V$. Then in particular, $\tSQMD(\hat V^+) = \MD(\hat S\cup \hat H^+\cap \hat Q)=p$. Note that voters moving from $\hat H$ are not restricted to $\{r,p\}$. However, in any profile $\hat H'$ where $\alpha|\hat H|$ voters change their vote, we still have a majority of honest voters on $r$, and thus $\MD(\hat H')=r$. This entails $\ol \MD_\alpha(\hat H) = \{r\}$ and thus 
    $$\tSQMD(\hat V^+) =p \notin \{r\} = \calB(r,\{r\}) = \calB(r,\ol \MD_\alpha(\hat H)\},$$
    which is a violation of $\safe$-safety on the tree or line.
    
\medskip
    In the other direction, suppose that  $\tSQMD$ is not  $\safe$-safe w.r.t. the median for a given set of parameters, and consider a  profile $V=H^+\cup H^-\cup S$ on a tree or a line where $\safe$-safety is violated. Then $p:=\tSQMD(V^+)=\MD(H^+\cup S\cup Q)$ and $p\notin \calB(r,\ol \MD_{\safe}(H))$. This means that for any `safe' position $b\in \calB(r,\ol \MD_{\safe}(H))$, and for a majority of voters $i\in H^+\cup S\cup Q$, $p$ is between $a_i$ and $b$. In particular this is true for the safe point $r'\in \calB(r,\ol \MD_{\safe}(H))$ that is closest to $p$. See an example in Fig.~\ref{fig:tree_reduction}. 
    
    We now construct a binary instance by considering only the two alternatives $\{p,r'\}$, with $r'$ in the role of status quo among the two. We project all voters onto their more preferred position among $p$ and $r'$. Then by Lemma~\ref{lemma:MD_MJ}, $\ol \MJ_\alpha(H|_{pr'})=\{r'\}$, since $r'$ is between $p$ and a super-majority of honest voters (6 out of 7 in Fig.~\ref{fig:tree_reduction}); whereas 
    $$\tSQMJ(V^+|_{pr'}) = p \notin \{r'\} = \calB(r',\{r'\}) = \calB(r',\ol \MJ_\alpha(H|_{pr'})), $$
    since $p$ is between $r'$ and most active voters (7 out of 13 in the example).
    We get that even in a binary setting the same set of parameters does not guarantee $\safe$-safety w.r.t. the majority rule.
\end{proof}

\begin{figure}
    \centering

\begin{tikzpicture}[
    scale=1.3,
    vertex/.style={circle, draw, minimum size=0.6cm, inner sep=0pt, fill=white},
    safe/.style={circle, draw, fill=green!30, minimum size=0.6cm, inner sep=0pt},
    special/.style={circle, draw, fill=green!14, minimum size=0.6cm, inner sep=0pt},
    edge/.style={thick},
    safe edge/.style={
          draw,
        line width=0.8pt,
        preaction={
            draw,
            green!70!black,
            dashed,
            line width=0.8pt,
            transform canvas={shift={(1.2pt,-1.2pt)}}
        }    
        },
    alpha edge L/.style={
         draw,
        line width=0.8pt,
        preaction={
            draw,
            green!70!black,
            line width=0.8pt,
            transform canvas={shift={(-1.2pt,-1.2pt)}}
        }
    },
    alpha edge R/.style={
         draw,
        line width=0.8pt,
        preaction={
            draw,
            green!70!black,
            line width=0.8pt,
            transform canvas={shift={(1.2pt,-1.2pt)}}
        }
    }
]

\tikzstyle{dot}=[rectangle,draw=black,fill=white,inner sep=0pt,minimum size=4mm]
\tikzstyle{active}=[circle,draw=blue,fill=blue,inner sep=0pt,minimum size=2mm]
\tikzstyle{inactive}=[circle,draw=blue,fill=white,inner sep=0pt,minimum size=2mm]
\tikzstyle{sybil}=[rectangle,draw=black,fill=red,inner sep=0pt,minimum size=1.6mm]
\tikzstyle{del}=[triangle,draw=blue,fill=white,inner sep=0pt,minimum size=2mm]
\tikzstyle{virtual}=[diamond,draw=black,fill=gray,inner sep=0pt,minimum size=2mm]
\tikzstyle{txt}=[text width = 8cm, anchor=west]

\node[safe] (C_b) at (0, 0) {$r'$};  
\node[safe] (LL) at (-2, 1) { };
\node[vertex] (CC_p) at (0, 1) {$p$};
\node[sybil] at (0.4,1) { };
\node[special] (RR) at (2, 1) { };  
\node[safe] (LLL) at (-3, 2) { };
\node[active]  at (-3,2) { }; 
\node[safe] (LLC) at (-2, 2) { };  
\node[active]  at (-2.1,2) { }; 
\node[inactive]  at (-1.9,2) { }; 
\node[vertex] (LLR) at (-1, 2) { };  
\node[vertex] (CCC) at (0, 2) { };  
\node[active]  at (0,2) { }; 
\node[sybil] at (0.4,2) { };
\node[sybil] at (0.6,2.01) { };
\node[sybil] at (0.8,2.02) { };
\node[vertex] (CCCC) at (0,3) { };
\node[sybil] at (0.4,3) { };
\node[sybil] at (0.6,3.02) { };
\node[special] (RRC_r) at (2, 2) {$r$};  
\node[virtual] at (2.4,2) { };
\node[virtual] at (2.6,2) { };
\node[vertex] (LLLL) at (-4, 3) { };
\node[vertex] (LLLC) at (-3, 3) { };  
\node[active]  at (-3,3) { }; 
\node[vertex] (RRCC) at (2, 3) { };  
\node[active]  at (2,3) { }; 
\node[safe] (L) at (-1,0.5) { };
\node[safe] (R) at (1,0.5) { };
\node[inactive]  at (1,0.5) { };

\draw[alpha edge L] (C_b) -- (L);
\draw[alpha edge L] (L) -- (LL);
\draw[edge] (C_b) -- (CC_p);
\draw[alpha edge R] (R) -- (C_b);
\draw[safe edge] (RR) -- (R);
\draw[alpha edge L] (LL) -- (LLL);
\draw[alpha edge L] (LL) -- (LLC);
\draw[edge] (LL) -- (LLR);
\draw[edge] (CC_p) -- (CCC);
\draw[edge] (CCCC) -- (CCC);
\draw[safe edge] (RRC_r) -- (RR);
\draw[edge] (LLL) -- (LLLL);
\draw[edge] (LLL) -- (LLLC);
\draw[edge] (RRC_r) -- (RRCC);

\draw (3.3,0) rectangle (5.5,2);
\node[vertex] (r_new) at (4, 0.5) {$r'$};  
\node[active]  at (4.4,0.6) { }; 
\node[active]  at (4.6,0.6) { }; 
\node[virtual]  at (4.8,0.6) { }; 
\node[virtual]  at (5,0.6) { }; 
\node[active]  at (4.4,0.4) { }; 
\node[active]  at (4.6,0.4) { }; 
\node[inactive]  at (4.8,0.4) { }; 
\node[inactive]  at (5,0.4) { }; 

\node[vertex] (p_new) at (4, 1.5) {$p$};
\node[active]  at (4.4,1.6) { }; 
\node[sybil]  at (4.6,1.6) { }; 
\node[sybil]  at (4.8,1.6) { }; 
\node[sybil]  at (5,1.6) { }; 
\node[sybil]  at (4.6,1.4) { }; 
\node[sybil]  at (4.8,1.4) { }; 
\node[sybil]  at (5,1.4) { }; 


\end{tikzpicture}
    \caption{An example of approximate safety violation on a tree, for $\alpha=\frac2{h}$. The sybils and virtual voters are drawn next to their position.  We mark  $\ol \MD_\alpha(H)$  with solid green (all nodes that can become the median of $H'$ by moving two of the seven honest agents). The remaining safe area between $ \MD_\alpha(H))$ and $r$ is marked by light/dashed green. The median of all active voters (honest, sybil, and virtual) is at $p$, whereas $r'$ is the nearest safe point. In general $r'$ and $p$ may not be adjacent.\newline The box on the right shows the induced binary instance.}
    \label{fig:tree_reduction}
\end{figure}

Just as in Section~\ref{sec:domains}, we get the safety properties of $\tSQMD$ as an immediate corollary from Theorems \ref{thm:SQMJ_safe_alpha} and \ref{thm:MJ_MD_reduction_alpha}:
\begin{corollary}
$\tSQMD$ is $\safe$-safe w.r.t $\MD$ as the base rule if and only if $\safe \geq\frac{1+\sigma-(1+\tau)(1-\mu)}{2(1-\sigma)}$.    
\end{corollary}
}
\paragraph{Multiple referenda}
The $\tREIMJ$ rule does not inherit the approximate-safety properties of $\tSQMJ$ for $\safe>0$. Intuitively, this is since honest voters might be split and only have weak agreement on each issue, which provides fewer sybils with enough power to thwart the decision.
\begin{proposition}\label{prop:IMJ_alpha_bad}
    For $\safe>0$, the $\safe$-safety guarantees of $\tREIMJ$ with respect to $\IMJ$ are strictly worse than those of $\tSQMJ$ with respect to $\MJ$. 

    This is true even with full participation ($\mu=0$) and without virtual voters ($\tau=0$).
\end{proposition}
\begin{proof} We show via an explicit example.

Suppose that $|H|=60, |S|=20$ (i.e. $\sigma = 1/4$),  $\tau=0, \mu=0$.
Then by Thm.~\ref{thm:SQMJ_safe_alpha} we get $1/6$-safety of the $\MJ$ rule with respect to itself  (indeed, if there are 40 honest voters on `0` and 20 on `1`, then moving $10 = |H|/6$ to `1' is sufficient).

Now consider $A=\{0,1\}^3$, with the status quo at $r=(0,0,0)$.  Honest voters are dispersed as follows:
20 on $(0,0,1)$; 
20 on $(0,1,0)$;
20 on $(1,0,0)$ and
all 21 voters of $S$ are on $(1,1,1)$ so the outcome is $\IMJ(V) = (1,1,1)$.

However we argue that $(1,1,1)\notin \calB(r,\ol{\IMJ}_\frac16(H))$ which means a violation of $\frac16$-safety.

Note that for this it is sufficient to show that there is no $H_\rightarrow$ with $|H\cap H'|\geq \frac56|H|=50$ s.t. $\IMJ(H') = (1,1,1)$.

Indeed, only 10 voters are allowed to vote differently in $H'$ than in $H$. Consider the original vote of an arbitrary `changed' voter $i$ in $H'\setminus H$. W.l.o.g.  $i$ voted $(1,0,0)$. This means there can be at most other 9 voters in $H'\setminus H$ whose original vote on the first issue is `0', and thus at most 9 new votes to `1' on the first issue.

Therefore, in $H'$ there are at least 31 votes to `$0$' vs. at most 29 votes to `$1$', meaning in particular that $\IMJ(H')\neq (1,1,1)$.

That is, $\IMJ$ is \emph{not} $1/6$-safe with respect to itself, in contrast to $\MJ$ with the same parameters $\sigma,\mu$ and $\tau$.
\end{proof}
By moving $5$ voters from each location to $(1,1,1)$, i.e. 15 in total, $\IMJ$ would select $(1,1,1)$. This entails that $\IMJ$ is  $\alpha$-safe with respect to itself  for $\alpha=\frac{15}{60}=\frac14$ (on the above profile), and it is not hard to see that this is tight. 

A similar example can be constructed for Condorcet-conservative rules, where different sets of honest voters prefer $p$ over $p'$ for each $p'$. 



\subsection{Quantifying Liveness}

It is possible to quantify liveness in a similar way, by requiring only that every outcome $p\in A$ is included in the \emph{outcome range} of the active voters when some fraction of up to $\beta$ of honest voters change their vote. Then we would get the standard definition of liveness for $\beta=1$, whereas lower values represent a \emph{stronger} livness requirement; and higher values than 1 represent a relaxed requirement. 

Since we see quantifying liveness as less natural and less interesting than quantified safety, we defer the technical details to Appendix~\ref{apx:liveness}.


\section{Random Participation}\label{sec:rand_finite}

The lower bound in Theorem~\ref{thm:arb_LB} suggests that no mechanism  can accommodate higher abstention and sybil rates than the $\tSQMJ$ mechanism, even in a binary setting. 
This, however, holds in the `worst case', making   adversarial assumptions both on the sybils' votes and on who chooses to abstain. 

A less extreme approach that might be more realistic is that the active honest voters are selected \emph{uniformly at random} from the honest population, whereas sybils  still vote adversarially.
%
%
%
%
%
As a result, we have that the votes of the active and the passive honest voters are similarly distributed. 

The benefit of such an assumption is demonstrated in Fig.~\ref{fig:sto}, where the `bad' selection of active voters on the left is possible under arbitrary participation, but highly unlikely under random participation. 

We argue that with this additional constraint on vote distributions, the safety-liveness tradeoff could be improved. However, since the  votes are now stochastic, the outcome is a random variable, and so we must first adapt our definitions, and in particular state what distribution of outcomes is considered `safe'. 

Alternatively, we can consider the limit case of a very large population, where the distributions of passive and active (honest) voters over alternatives are \emph{exactly} the same, as any variance becomes negligible. This `nonatomic' model is somewhat easier to analyze, but yields similar results and is deferred to Appendix~\ref{sec:rand_nonatomic}.

In the remainder of this section we consider finite populations. This requires a probabilistic extension of the safety and liveness properties.

\subsection{Safety for Stochastic Outcomes}
\begin{figure}
    \centering
\begin{tikzpicture}[scale=0.4]

\tikzstyle{dot}=[rectangle,draw=black,fill=white,inner sep=0pt,minimum size=4mm]
\tikzstyle{active}=[circle,draw=blue,text=white,fill=blue,inner sep=0pt,minimum size=3mm]
\tikzstyle{inactive}=[circle,draw=blue,fill=white,inner sep=0pt,minimum size=3mm]
\tikzstyle{sybil}=[rectangle,draw=black,fill=red,text=white,inner sep=0pt,minimum size=3mm]
\tikzstyle{del}=[triangle,draw=blue,fill=white,inner sep=0pt,minimum size=2mm]
\tikzstyle{virtual}=[diamond,draw=black,fill=gray,inner sep=0pt,minimum size=4mm]
\tikzstyle{txt}=[text width = 2cm, anchor=west]

\node at (2,6) {\textit{Skewed sample}};

\draw (-1,0.4) -- (8,0.4);
\node at (1,-.2) {$r$};
\node at (6,-.2) {$p$};


\node at (1,2) [active] {};
\node at (2,2) [inactive] {};
\node at (3,2) [inactive] {};
\node at (1,1) [inactive] {};
\node at (2,1) [inactive] {};
\node at (3,1) [inactive] {};
\node at (1,4) [virtual] {};
\node at (2,4) [virtual] {};

\node at (6,3) [sybil] {};
\node at (7,3) [sybil] {};

\node at (6,2) [active] {};
\node at (7,2) [active] {};
\node at (6,1) [active] {};
\node at (7,1) [active] {};


\end{tikzpicture}
~~~~~~~~~~
\begin{tikzpicture}[scale=0.4]

\tikzstyle{dot}=[rectangle,draw=black,fill=white,inner sep=0pt,minimum size=4mm]
\tikzstyle{active}=[circle,draw=blue,text=white,fill=blue,inner sep=0pt,minimum size=3mm]
\tikzstyle{inactive}=[circle,draw=blue,fill=white,inner sep=0pt,minimum size=3mm]
\tikzstyle{sybil}=[rectangle,draw=black,fill=red,text=white,inner sep=0pt,minimum size=3mm]
\tikzstyle{del}=[triangle,draw=blue,fill=white,inner sep=0pt,minimum size=2mm]
\tikzstyle{virtual}=[diamond,draw=black,fill=gray,inner sep=0pt,minimum size=4mm]
\tikzstyle{txt}=[text width = 2cm, anchor=west]

\node at (2,6) {\textit{Common sample}};

\draw (-1,0.4) -- (8,0.4);
\node at (1,-.2) {$r$};
\node at (6,-.2) {$p$};


\node at (1,2) [active] {};
\node at (2,2) [active] {};
\node at (3,2) [active] {};
\node at (1,1) [inactive] {};
\node at (2,1) [inactive] {};
\node at (3,1) [inactive] {};
\node at (1,4) [virtual] {};
\node at (2,4) [virtual] {};

\node at (6,3) [sybil] {};
\node at (7,3) [sybil] {};

\node at (6,2) [active] {};
\node at (7,2) [active] {};
\node at (6,1) [inactive] {};
\node at (7,1) [inactive] {};


\end{tikzpicture}
\caption{\label{fig:sto}Two possible realizations of the same instance with $|H_p|=4$, $|H_r|=6$ and the same number of active honest voters  $n^+=5$. In the realization on the left, most active voters are on $p$ and thus $p$ wins (violating safety). On the realization on the right, exactly half of the honest voters on each alternative are active, and thus the only safe alternative $r$ wins.}
\end{figure}

\revision{
\paragraph{Probabilistic safety}
We next extend the notion of safety to settings involving randomness. 
Randomness may arise from voters’ behavior (e.g., deciding probabilistically whether to participate), 
from the aggregation rule~$\mathcal{R}$ itself, or from other external sources. 
The base rule~$\mathcal{G}$, however, is assumed to remain deterministic. 

\medskip
\noindent
Let $n^{+} := |H^{+}| = |V| \cdot h^{+}$ denote the number of active honest voters. 
Safety \emph{with high probability} means that when this number is sufficiently large, 
the likelihood of obtaining an unsafe outcome becomes negligible.\footnote{
For a given instance, we treat the number of active voters as fixed, meaning they are selected from the honest population without repetition. 
One could alternatively assume that each honest voter is active with some fixed probability; 
the results would be similar, though the definitions of both safety and liveness would require minor adjustments.}
We highlight in {\color{blue}{blue}} the differences from Definition~\ref{def:safe_quant}.

\medskip
\noindent
In this probabilistic setting, an instance~$V$ specifies only the partition into honest and sybil voters, but not which honest voters are active.

\begin{definition}[Safety w.h.p.]\label{def:safe_whp}
An aggregation rule $\mathcal{R}$ is $\alpha$-\emph{safe {\color{blue!50!black}{with high probability}}} with respect to $\mathcal{G}$ if 
{\color{blue}{for any $\alpha' > \alpha$, there exists a constant $C$ such that for all populations $V$ with $n^{+}$ active voters,}}
\[
{\color{blue!50!black}{
\Pr_{x \sim \mathcal{R}(V)}[}}x \in \mathcal{B}(r; \overline{\mathcal{G}}_{\alpha'}(H))
{{\color{blue!50!black}{] > 1 - \exp(-C \cdot n^{+})}.
}}
\]
\end{definition}

\noindent
In words, as the number of active honest voters grows, the probability that the mechanism produces an unsafe outcome decays exponentially. Note that probability is taken over the random selection of $H^+$ (uniform without replacement), and any internal randomness of $\calR$, if there is any. 
This formalizes the idea that safety holds \emph{with overwhelming probability} in large electorates.
}

The constant $C$ may depend on the instance parameters ($\sigma,\mu,\alpha$)  and as specified also on  $\alpha'$.  Note that the requirement of safety w.h.p. is no longer for a given instance (as it is asymptotic), but on all instances with given parameters. 

We could similarly define liveness w.h.p., and this would make sense for various sources of uncertainty, but for our particular model this is not required:
since there are exactly $(1-\mu-\sigma)|V|$ active honest voters, and since in the worst case for liveness, all voters vote for $r$, all realizations are identical. The probability that there is a violation of liveness is thus either $0$ or $1$. 



\subsection{The Binary Case}
We show an improved bound compared to the arbitrary participation case (Thm.~\ref{thm:SMJ_safe_alpha}).
\begin{theorem}\label{thm:SQMJ_random_alpha}
Under random participation,    the $\tSQMJ$ voting rule is $\safe$-safe w.h.p. with respect to Majority, iff $\safe \geq \frac{(\sigma - \tau(1 - \mu))(1-\sigma)}{2(1 - \mu - \sigma)}$. 
\end{theorem}

\begin{proof}
Recall we denote by $u_r,u_p$ the fraction of voters for $r$ and $p$, respectively, in a voter set $U$. 

Consider any $\safe'>\safe$. In the case where $h_p > h_r -2\safe'$, we have
$$p\in \ol{MJ}_{\safe'}(H)\subseteq  \calB(r;\ol{MJ}_{\safe'}(H)),$$
which means $\safe$-safety holds regardless of the realization of active voters.

\medskip

Therefore, assume that $h_p \leq h_r -2\safe'$.
Intuitively, this means that the gap $h_r-h_p$ is large, and thus the gap $h^+_r-h_p^+$  is likely to be large as well, leading to $v^+_r > v^+_p$ w.h.p. 
E.g. in the `common' realization on Fig.~\ref{fig:sto}, we have $h^+_r-h_p^+ = \frac{3}{12}-\frac{2}{12}=\frac{1}{12}$ (Right figure).

 This is the main difference from the arbitrary participation case where must also  consider highly skewed realizations (E.g. in the Left of Fig.~\ref{fig:sto} the gap is $-\frac14$ and $v^+_p$ is indeed strictly higher than $v^+_r$).
 
The remainder of the proof is for showing, using the Hoeffding inequality, that w.h.p the  gap $h^+_r-h^+_p$ is larger than $s-q$, and hence $r$ has more active votes overall, and safety is not violated. We now turn to prove this formally.




\medskip  
To show safety w.h.p., we need to upper-bound the probability  that $\tSQMJ$ will select $p$. 

Denote $c:=\safe'-\safe>0$.
Since $h_p \leq h_r -2\safe'$, and by the premise of the theorem, we have:
\labeq{beta_rand}
{\frac{(\sigma - \tau(1 - \mu))(1-\sigma)}{2(1 - \mu - \sigma)} +c \leq \safe+c=\safe' \leq \frac{h_r-h_p}{2}.}

 Every honest voter is active with probability $\phi:=\frac{|H^+|}{|H|}=\frac{1-\mu-\sigma}{1-\sigma}$ (though not i.i.d). 
Alternatively, every active honest voter is a $p$ voter with probability $\psi:=|H_p|/|H|$.

We sample $n^+=|H^+|$ active voters from the set $H=H_p \cup H_r$, \emph{without replacement}. Consider $n^+$ samples $X_1,\ldots,X_{n^+}\in \{0,1\}$ where $X_i=1$ if the $i$'th active agent is a $p$ voter, and $0$ otherwise. Thus $n^+_p:=|H^+_p|=\sum_{i\leq n^+}X_i$ and  $n^+_r=n^+-n^+_p$. 

Observe that $n^+_p$ is a random variable, whose expected value is 
$$n^+\cdot E[X_i] = n^+ \psi =|H^+| \frac{|H_p|}{|H|}= \frac{|H^+|}{|H|}|H_p|=\phi |H_p|.$$

Recall that $c=\safe'-\safe$ and let $\eps\in (0,\frac{c}{2-2\sigma})$.
Denote the event $[n^+_p< (\psi+\eps)n^+]$ by $I$.
By applying Hoeffding inequality,\footnote{The Hoeffding inequality applies for sampling either with or without replacement. Without replacement it is possible to get somewhat better bounds~\cite{serfling1974probability} but this is immaterial for our argument.} 
$$Pr[\neg I]=Pr[n^+_p \geq (\psi+\eps)n^+] < exp(-2\eps^2 n^+) = exp(-c^2 n^+ / (1-\sigma)^2)= exp(-C\cdot n^+),$$
for $C=\left(\frac{\safe'-\safe}{1-\sigma}\right)^2$. 
It thus remains to show that whenever $I$ occurs, $r$ is selected. 

\medskip
For the remainder of the proof, we fix a realization where event $I$ occurs, thus $n^+_p< (\psi+\eps)n^+= \phi |H_p| + \eps n^+$, and $n^+_r = n^+-n^+_p > (1-\psi-\eps)n^+ = \phi |H_r| -\eps n^+$ (intuitively, $n^+_p,n^+_r$ are close to their expected values). 
Therefore:
  \labeq{eps_rand}
  {h^+_r-h^+_p = \frac1n(n^+_r-n^+_p) > \frac1n(\phi(|H_r|-|H_p|) - 2\eps n^+) = \phi(h_r-h_p) -2\eps (1-\sigma-\mu). 
}

By definition,  $\tSQMJ(V^+) = \MJ(H^+ \cup S \cup Q)$ where $Q$ contains $\tau|V^+| = \tau(1-\mu)n$ voters for $r$.

Thus the total fraction of active $r$ voters is at least $h^+_r+\tau(1-\mu)$. As in the previous proofs, w.l.o.g. all sybils vote for $p$ as this is the worst case for safety. We get that
\begin{align*}
v^+_r&-v^+_p \geq (h^+_r + \tau(1-\mu)) - (h^+_p + \sigma) \\
&=   (h^+_r-h^+_p) - (\sigma- \tau(1-\mu))\\
&>    \phi(h_r-h_p)-2\eps(1-\sigma-\mu)  - (\sigma- \tau(1-\mu)) \tag{by Eq.~\eqref{eq:eps_rand}}\\
&> \frac{1-\mu-\sigma}{1-\sigma}\left( \frac{(\sigma - \tau(1 - \mu))(1-\sigma)}{1 - \mu - \sigma} + 2c\right) \\
&~~-2\eps(1-\sigma-\mu)- (\sigma- \tau(1-\mu))\tag{by Eq.~\eqref{eq:beta_rand}}\\
&= \frac{1-\mu-\sigma}{1-\sigma}\cdot\frac{(\sigma - \tau(1 - \mu))(1-\sigma)}{1 - \mu - \sigma}\\
&~~+ 2(1-\mu-\sigma)(\frac{c}{1-\sigma}-\eps) - (\sigma- \tau(1-\mu))\\
&=  (\sigma- \tau(1-\mu))  + 2(1-\mu-\sigma)(\frac{c}{1-\sigma}-\eps) - (\sigma- \tau(1-\mu))\\
&>0 \tag{since $\eps<\frac{c}{1-\sigma}$},
\end{align*}
as required. 

\medskip Tightness follows from the same construction used in the nonatomic case. Then there are strictly more active voters (in expectation) for the unsafe alternative $p$, and the probability of selecting $p$ is at least $0.5$.
\end{proof}

We can therefore trace the improved tradeoff between safety and liveness as follows:

\begin{corollary}\label{cor:SQMJ_random}
  Under  random participation, the following holds:
  \begin{itemize}
      \item $\tSQMJ$ is $0$-safe w.h.p w.r.t $\MJ$ iff $\tau \geq \frac{\sigma}{1 - \mu}$.
    \item $\tSQMJ$ is live iff $\tau< \frac{1-2\sigma-\mu}{1-\mu}$.
   Which provides us with the valid interval:
  \end{itemize}
  \begin{center}\boxed{\frac{\sigma}{1 - \mu} \leq \tau < \frac{1-2\sigma-\mu}{1-\mu}~~~~\textit{  iff  }~~~~~3\sigma + \mu < 1}\end{center}
\end{corollary}
This is compared to $3\sigma+2\mu<1$ requirement in the arbitrary participation model (Thm.~\ref{thm:SQMJ}). Therefore, adversarial abstention is `twice as bad' as random abstention.



\subsection{Extensions Beyond the Binary Case}

Note that our definition for `safety w.h.p' is general and applies to any domain. 

All of our positive results use reductions to the binary case: either to Thm.~\ref{thm:SQMJ} (if restricted to $\safe=0$); or to Thm.~\ref{thm:SQMJ_safe_alpha} (when apply to any $\safe\geq 0$). The same reductions would apply for the random participation model, using Cor.~\ref{cor:SQMJ_random} or Thm.~\ref{thm:SQMJ_random_alpha}, respectively. 

Thus all of our previous results extend to the random participation model, with the improved bound. This applies to:
\begin{itemize}
    \item Multiple alternatives (Thm.~\ref{thm:SMJ_safe}, Thm.~\ref{thm:SMJ_safe_alpha}, Prop.~\ref{prop:IMJ}, Prop.~\ref{prop:SCC});
    \item Multiple referenda (Prop.~\ref{prop:IMJ});
    \item Single-peaked domains (Thm.~\ref{thm:MJ_MD_reduction}, Thm.~\ref{thm:MJ_MD_reduction_alpha}).
\end{itemize}

\section{Voting with Delegation}\label{sec:delegation}

While the results above allow for partial participation, they also imply that to obtain both safety and liveness, the fraction of passive voters cannot be too large;
this might be problematic in some situations. As our lower bound means that this is unavoidable, we therefore wish to relax the model to analyze other possibilities; in particular, we adopt the standard model of \emph{proxy voting}, where only a small number of voters are active, and any passive voter delegates her vote to the nearest active voter~\cite{alger2006voting,cohensius2017proxy}.
\short{\footnote{We note that in the binary domain, proxy voting trivially yields the same bounds as in Cor.~\ref{cor:REMJ}, as long as $r$ has at least one voter.}}
%
 
\paragraph{Voting with a constant number of alternatives}
There is no reason in doing a separate analysis for delegation in the binary (or any categorical) domain, as, in this domain there is no difference between delegating to a proxy and actively voting (provided that every alternative has at least one active voter); sybils may still interfere, but the safety-liveness tradeoff of Majority with proxy delegation is just as in Thm.~\ref{thm:SQMJ} with full participation ($\mu=0$).

In contrast, in continuous or structured domains, an inactive voter will  rarely find an active voter that completely agrees, and thus the effect of delegation becomes nontrivial. 

%
%




\subsection{Median with Delegation on a Line}
\begin{figure}[t]
\begin{center}
\begin{small}

\begin{tikzpicture}[scale=0.6]

\tikzstyle{dot}=[rectangle,draw=black,fill=white,inner sep=0pt,minimum size=4mm]
\tikzstyle{active}=[circle,draw=blue,fill=blue,inner sep=0pt,minimum size=2mm]
\tikzstyle{inactive}=[circle,draw=blue,fill=white,inner sep=0pt,minimum size=2mm]
\tikzstyle{sybil}=[rectangle,draw=black,fill=red,inner sep=0pt,minimum size=1.6mm]
\tikzstyle{del}=[triangle,draw=blue,fill=white,inner sep=0pt,minimum size=2mm]
\tikzstyle{virtual}=[diamond,draw=black,fill=gray,inner sep=0pt,minimum size=2mm]
\tikzstyle{txt}=[text width = 8cm, anchor=west]

\node at (0,2) [txt] {(a) $V=H^+\cup H^- \cup S$};
\draw (0,0.5) -- (20,0.5);
\draw (4,0.3) -- (4,0.7);
\node at (4,0) {$r$};
\node at (0,-0.5) {};

\node at (2,1) [inactive] {};
\node at (7,1) [inactive] {};
\node at (11,1) [inactive] {};
\node at (12,1) [inactive] {};
\node at (12,1.4) [inactive] {};
\node at (17,1) [inactive] {};
\node at (16,1) [inactive] {};
\node at (5,1) [active] {};
\node at (13,1) [active] {};
\node at (8,1) [sybil] {};
\node at (15,1) [sybil] {};
\node at (15,1.4) [sybil] {};

%
%
%
%
%

\end{tikzpicture}

\begin{tikzpicture}[scale=0.6]

\tikzstyle{dot}=[rectangle,draw=black,fill=white,inner sep=0pt,minimum size=4mm]
\tikzstyle{active}=[circle,draw=blue,fill=blue,inner sep=0pt,minimum size=2mm]
\tikzstyle{inactive}=[circle,draw=blue,fill=white,inner sep=0pt,minimum size=2mm]
\tikzstyle{sybil}=[rectangle,draw=red,fill=red,inner sep=0pt,minimum size=2mm]
\tikzstyle{del}=[triangle,draw=blue,fill=white,inner sep=0pt,minimum size=2mm]
\tikzstyle{virtual}=[diamond,draw=black,fill=gray,inner sep=0pt,minimum size=2mm]
\tikzstyle{txt}=[text width = 8cm, anchor=west]

\node at (0,-0.5) {};
\node at (0,2) [txt] {(b) $z^*=MD(H)$};
\draw (0,0.5) -- (20,0.5);
\draw (4,0.3) -- (4,0.7);
\node at (4,0) {$r$};
\draw (12,0.3) -- (12,0.7);
\node at (12,0) {$z^*$};
\draw (13,0.3) -- (13,0.7);
\node at (2,1) [inactive] {};
\node at (7,1) [inactive] {};
\node at (11,1) [inactive] {};
\node at (12,1) [inactive] {};
\node at (12,1.4) [inactive] {};
\node at (17,1) [inactive] {};
\node at (16,1) [inactive] {};
\node at (5,1) [active] {};
\node at (13,1) [active] {};

%
%
%
%
%

\end{tikzpicture}

\begin{tikzpicture}[scale=0.6]

\tikzstyle{dot}=[rectangle,draw=black,fill=white,inner sep=0pt,minimum size=4mm]
\tikzstyle{active}=[circle,draw=blue,fill=blue,inner sep=0pt,minimum size=2mm]
\tikzstyle{inactive}=[circle,draw=blue,fill=white,inner sep=0pt,minimum size=2mm]
\tikzstyle{sybil}=[rectangle,draw=black,fill=red,inner sep=0pt,minimum size=1.6mm]
\tikzstyle{del}=[triangle,draw=blue,fill=white,inner sep=0pt,minimum size=2mm]
\tikzstyle{virtual}=[diamond,draw=black,fill=gray,inner sep=0pt,minimum size=2mm]
\tikzstyle{txt}=[text width = 8cm, anchor=west]

\node at (0,-0.5) {};
\node at (0,2.5) [txt] {(c) $\hat h=\tREMD(V)=\MD(V \cup Q)$};
\draw (0,0.5) -- (20,0.5);
\draw (4,0.3) -- (4,0.7);
\node at (4,0) {$r$};
\draw (11,0.3) -- (11,0.7);
\node at (11,0) {$y$};
\node at (4,1) [virtual] {};
\node at (4,1.3) [virtual] {};
\node at (4,1.6) [virtual] {};
\node at (2,1) [inactive] {};
\node at (7,1) [inactive] {};
\node at (11,1) [inactive] {};
\node at (12,1) [inactive] {};
\node at (12,1.4) [inactive] {};
\node at (17,1) [inactive] {};
\node at (16,1) [inactive] {};
\node at (5,1) [active] {};
\node at (13,1) [active] {};
\node at (8,1) [sybil] {};
\node at (15,1) [sybil] {};
\node at (15,1.4) [sybil] {};

%
%
%
%
%

\end{tikzpicture}

\begin{tikzpicture}[scale=0.6]

\tikzstyle{dot}=[rectangle,draw=black,fill=white,inner sep=0pt,minimum size=4mm]
\tikzstyle{active}=[circle,draw=blue,fill=blue,inner sep=0pt,minimum size=2mm]
\tikzstyle{inactive}=[circle,draw=blue,fill=white,inner sep=0pt,minimum size=2mm]
\tikzstyle{sybil}=[rectangle,draw=black,fill=red,inner sep=0pt,minimum size=1.6mm]
\tikzstyle{del}=[diamond,draw=blue,fill=white,inner sep=0pt,minimum size=2mm]
\tikzstyle{virtual}=[diamond,draw=black,fill=gray,inner sep=0pt,minimum size=2mm]
\tikzstyle{txt}=[text width = 8cm, anchor=west]

\node at (0,2.8) [txt] {(d) $x=\tREMD^P(V^+;\vec w)=\MD^P(H^+ \cup S \cup Q;\vec w)$};
\draw (0,0.5) -- (20,0.5);
\draw (4,0.3) -- (4,0.7);
\node at (4,0) {$r$};
\draw (13,0.3) -- (13,0.7);
\node at (13,0) {$x$};

\node at (4,1) [virtual] {};
\node at (4,1.3) [virtual] {};
\node at (4,1.6) [virtual] {};
\node at (5,1) [active] {};
\node at (13,1) [active] {};
\node at (8,1) [sybil] {};
\node at (15,1) [sybil] {};
\node at (15,1.4) [sybil] {};
\node at (4,1.9) [del] {};
\node at (8,1.4) [del] {};
\node at (13,1.3) [del] {};
\node at (13,1.6) [del] {};
\node at (13,1.9) [del] {};
\node at (15,1.6) [del] {};
\node at (15,1.9) [del] {};

\draw[dotted] (14,0.8) -- (14,2);
\draw[dotted] (10.5,0.8) -- (10.5,2);
\draw[dotted] (6.5,0.8) -- (6.5,2);
\draw[dotted] (4.5,0.8) -- (4.5,2);
%
%
%
%
%

\end{tikzpicture}
\end{small}
\end{center}
\caption{\label{fig:med}A demonstration of several definitions used in the proof of Theorem~\ref{thm:MD_del_safe} on an example profile. The honest voters are blue circles (filled circles are active voters $H^+$). Sybils are marked by red squares. The full gray diamonds are the virtual voters $Q$ added by the mechanism. In the bottom figure, hollow blue diamonds mark the followers of each active voter (their real positions are as in Fig.~(c), and the Voronoi partition is marked by dotted lines).}
\end{figure}
For a finite population $U$ and a vector of vote weights $\vec w=(w_i)_{i\in U}$, we denote by $MD(U;\vec w)$ the \emph{weighted median}, where each $i\in U$ has weight $w_i\in \mathbb N$. Formally, 
$$MD(U;\vec w):=\min\{u_i: i\in U, \sum_{j\leq i}w_j \geq \sum_{j>i}w_j\}.$$

Following Section~\ref{sec:rand_finite} we denote $n^+ := |H^+|\geq (1-\mu-\sigma)|V|$, and  assume that active voters are sampled uniformly at random from $H$.  As we will see later, the fraction of active voters itself will not matter and can be arbitrarily close to $0$. 

The votes of inactive voters affect the outcome indirectly via delegation:
  for each $i\in V^+$, let $w_i=1+|\{j\in V^- : i =\argmin_{i'\in V^+}|s_i-s_j|\}|$ be the number of voters for which $i$ is the closest active voter (their ``proxy'').
Indeed, this follows from our strong assumption, namely  that passive votes are always delegated to the closest active voter (either honest or sybil). We leave the study of alternative delegation models for future research. 

The rule $\MD^P$ ($P$ for Proxy) takes population $V=H\cup S$ as input together with the implicit parameter $n^+$, samples $n^+$ active voters from $H$, and returns $\MD(V^+;\vec w)$, where weights are set as above, according to the number of ``followers'' (i.e., delegatees) of each $i\in V^+ \cup \{r\}$.
Since $V^+=S\cup H^+$ and $\vec w$ are random variables, so is $\MD^P(V)$.

The rule $\tSQMD^P$ is the same, except adding $\tau|V|$ virtual voters on $r$ first, i.e. 
$$\tSQMD^P(V) = \MD^P(V\cup Q).$$


%

\full{
\begin{remark}\label{rem:r_active}
If, for some passive voter $i$, the status quo $r$ is closer than all active voters, then we assume that $i$ delegates to $r$ (see, e.g., Fig.~\ref{fig:med}(d)).
\end{remark}
}

\mypara{Analysis}
%
%
%

\begin{theorem}\label{thm:MD_del_safe}
    Under random participation, $\tSQMD^P$ is safe w.h.p. if and only if $\tau\geq \sigma$.
\end{theorem}
Let us use the following notation:
\begin{itemize}
    \item $X:=\tSQMD^P(V)$ is the returned position (which is a random variable);
    \item $z^* := \MD(H)$ is the honest outcome. We assume w.l.o.g. that $z^*\geq r$, so that the 0-safe range is $[r,z^*]$.
    \item $y:= \tSQMD(V)$, i.e. the median with sybils and virtual voters, but with full participation.
\end{itemize}
Note that $z^*$ and $y$ are fixed positions that do not depend on realization. 

In addition, we define by $z^-$ and $z^+$, respectively, the ends of the closed interval $\ol{MD}_{\alpha'}(H)$. Thus the $\safe'$-safe range is $[r,z^+]$. Still, $z^-,z^+$ are fixed positions. 

\begin{lemma}[Cohensius et al.~\citep{cohensius2017proxy}]\label{lemma:proxy}
  For any  $U=(U^+,U^-)$, it holds that $\MD(U^+;\vec w)$ with proxy weights is the voter in $U^+$ which is closest to $\MD(U)$. 
\end{lemma}

Our argument is as follows: we show that $y\leq z^* \leq z^+$, then use the lemma to argue that in every realization $x$ of $X$, the selected $x$ is the active voter closest to $y$. Finally, we show that w.h.p. there is some active voter in $[y,z^+]$ and thus $x\leq z^+$.

\begin{proof}[Proof of Theorem~\ref{thm:MD_del_safe}]
  By the premise of the theorem, $\tau\geq \sigma$. Since $y=\tSQMD(V)$ corresponds to an instance with full participation, we get from Cor.~\ref{cor:MD_tradeoff} with $\tau\geq \sigma$ and $\mu=0$ that $\tSQMD(V)$ is safe. Thus $r\leq y\leq z^*$.  

  Now consider Lemma~\ref{lemma:proxy}, where $U:= Q\cup H^+ \cup H^- \cup S$ is any realized partition of $V$ into active and inactive voters (with the added virtual voters $Q$).  We get that the realized outcome $x=\MD(U^+;\vec w)$ is the position of the voter in $U^+= Q\cup H^+\cup S$ which is closest to $\MD(U)=\tSQMD(V) = y$. In other words, if $i^*=\argmin_{i\in H^+ \cup S \cup Q}|s_i-y|$ is the closest active voter to $y$ (in some realization), then $x=s_{i^*}$.

   Since the virtual voters are active, we know $x\geq r$. It is left to show that with high probability there is an active voter between $y$ and $z^+$:
    We consider $n^+>2/\safe'$.
   
   indeed, the range $\ol{MD}_{\safe'}(H)$ contains $\floor{\safe'|H|}$ honest voters to each side of $z^*=\MD(H)$. 
   Since 
   \begin{align*}
       \floor{\safe'|H|} &\geq \safe'|H|-1= \frac12\safe'|H|+\frac12\safe'|H|-1\\
       &\geq \frac12\safe'|H|+\frac12\safe'n^+-1>\frac12\safe'|H|+\frac12\cdot 2-1\\
       &=\frac12\safe'|H|,
   \end{align*} 
   there are at least $\frac12\safe'|H|$ voters in $[z^*,z^+]$. Denote these voters by $\hat H$.
   Now, $H^+$ is a random sample of $n^+$ voters from $H$, so each voters $i\in H^+$ has a probability of at most $1-\frac12\safe'$ to be outside $\hat H$. Since we sample without repetition, by Hoeffding inequality the probability that all active voters are outside (i.e. that $H^+\cap \hat H$ is empty) is at most $(1-\frac12\safe')^{n^+} = \exp(-C\cdot n^+)$ for some positive constant $C$ that depends only on $\safe'$.
   Finally, 
   \begin{align*}
   \Pr_{x\sim X}[x\in \calB(r;\ol{MD}_{\alpha'}(H))] &= \Pr_{x\sim X}\left[x\in [r,z^+]\right]\\ &\geq Pr\left[H^+ \cap \hat H\neq\emptyset\right] 
   > 1-\exp(-C\cdot n^+),
   \end{align*}
   as required.
   
   \medskip
   In the other direction, if $\tau<\sigma$ then consider profiles where all voters are either on $r$ or on some other point $p$. By Cor.~\ref{cor:MD_tradeoff} this is unsafe even with full participation, i.e. there is an instance where most honest voters are on $r$ and yet $p$ is selected, meaning a majority (with some constant margin $\eps$) of voters from $V\cup Q$ are on $p$. Set $\alpha':=\eps/2$, then $p\notin \ol{\MD}_{\alpha'}(H)$. The probability that $p$ still wins when we sample the active voters is at least $1/2$ regardless of $n^+$, which means a violation of safety w.h.p.
\end{proof}

Delegation does not affect liveness: the $\tSQMD^P$ is live iff $\tau<1-2\sigma$, as this follows from the full participation case of Cor.~\ref{cor:MD_tradeoff}.

\begin{corollary}
  By setting $\tau=\sigma$, the $\tSQMD^P$ mechanism  is both safe w.h.p. and live, as long as $\sigma<\frac13$.
\end{corollary}
This shows that delegation allows us to almost completely eliminate the drawbacks of partial participation, and get the same safety level against sybils as with full participation, provided that the \emph{number} of active voters is sufficiently large (but without any requirement on their fraction).

\sect{Discussion and Outlook}\label{section:discussion}

We have analyzed different social choice settings in which sybil entities have infiltrated the voting community and, on top of this, not all honest voters participate. We have provided a formal model to reason about such situations, developed techniques to tackle this challenge, and analyzed them.

In particular, motivated by governance and mutual decision mechanisms for online communities, we have considered the common situation in which representation is threatened both by the presence of sybils, and by partial participation of the honest voters.
We have defined a general mechanism, $\tRER$, and analyzed its safety/liveness tradeoff for several social choice settings.
 For a fraction $\sigma$ of sybils and a fraction $\mu$ of passives in the population, we showed that, for voting on one proposal against the status quo and voting in an interval domain, the SQE mechanism can obtain maximal safety and liveness together as long as  $3\sigma+2\mu < 1$. 
Furthermore, we showed: that the same tradeoff applies to categorical decisions and to multiple referenda; that no mechanism can do better than $\tSQMJ$;  that we can be satisfied with a somewhat lower participation rate ($3\sigma+\mu < 1$) when participation is random; and that delegation allows the same level of safety with a negligible fraction of active honest voters.   


To set the parameter $\tau$ (the bias towards the status quo) effectively, after deciding upon the  desired tradeoff of safety and liveness, one has to estimate $\sigma$ and $\mu$ in the population.
While $\mu$ can be estimated quite accurately (as an election organizer may define the set of eligible voters), this is not the case for $\sigma$. The fraction of sybils can be approximated by sampling voters (see Remark~\ref{remark:sampling}) or by techniques that upper bound $\sigma$~\cite{csr}. Note that over-estimating $\sigma$ or $\mu$ always results in a mechanism that is more safe, and thus our bounds still hold.

Together with state-of-the-art mechanisms for identifying and eliminating sybils~\cite{sybilsurvey}, our results set the foundation for reliable and practical online governance tools.  Note also that, since the preliminary, conference version of this paper was published, it was identified as a crucial piece in the design of a democratic metaverse~\cite{shapirofoundations}.

Before we discuss some avenues for future research, we wish to comment on the practicality of our methods in the context of the estimation of the different parameters.

\paragraph{Estimating the sybil fraction}\label{remark:sampling}
How to estimate the sybil penetration $\sigma$ is an important question. While in some cases there might be other techniques available (some works on this topic -- including such in which $\sigma$ can theoretically be upper-bounded -- exist~\cite{csr,poupko2021building}), usually it is natural to assume that by sampling a voter one can estimate the probability that the voter is genuine or fake (e.g., looking at her Facebook profile). Thus, the main general technique we suggest is to sample voters uniformly at random and, given the sampling results, estimate $\sigma$. Note that using such sampling it is then possible to compute, for a given value $p$, a value $z$, such that the probability that $\sigma$ is greater than $z$ is at most $p$. Alternatively, one can compute the mean $m$ of the sample and take an $\epsilon$ margin of safety, i.e., use $m + \epsilon$ as the estimate for $\sigma$.

Finally, below we discuss several avenues for future research:
\begin{itemize}

\item
\textbf{Further social choice settings}:
In particular, generalizing some of our results to general metric spaces seems natural. In this context, we conjecture that $\tRER$, when applied to other metric spaces (with suitable base rules), would guarantee similar safety/liveness tradeoffs. 

\item
\textbf{Further delegation models}:
Relaxing the proxy voting assumption of Cohensius et al.~\shortcite{cohensius2017proxy} is a natural direction. In particular, considering more general and realistic delegation models that relate to some underlying social network and take into account voter affinity  seems promising.

\item
\textbf{Practical considerations}:
We feel that our theoretical framework and results are quite ready for being applied in the wild. However, to do so one may first go through performing extensive simulations, and then developing practical tools for communities to utilize the results presented here in a user-friendly, convenient, and robust way.

\end{itemize}





\forarxiv{
\section*{Acknowledgements}
We thank the generous support of the Braginsky Center for the Interface between Science and the Humanities.
Nimrod Talmon was supported by the Israel Science Foundation (ISF; Grant No. 630/19).
Reshef Meir is supported by the Israel Science Foundation (ISF; Grant No. 2539/20).
}

\bibliography{bib-journal}

\clearpage
\appendix
\onecolumn
\section{Quantifying Liveness}\label{apx:liveness}
Recall the original definition of liveness (Def.~\ref{def:live_general}), stating that $\calR$ is live w.r.t. $V=S \cup H$ if $\calR(S\cup H^+_{\rightarrow a}) = a$  for all $a\in A$.

We now relax this definition with a parameter $\live$.
\begin{definition}[$\live$-Liveness]\label{def:live_beta}
  An aggregation rule $\calR$ is $\live$-\emph{live} w.r.t. population $V$, if for all $a\in A$, it holds that $a\in \ol \calR_\live(V)$.  
\end{definition}

I.e., a rule is live w.r.t some population if any outcome can be reached by modifying not-too-many (in particular, $\live$-fraction of) honest voters.

For any monotone rule, $1$-liveness coincides with liveness. To see why, note that for $a$ to belong in the outcome range $\ol \calR_1(V)$, there must be \emph{some} honest profile $H'$ (with same size as $H$) s.t. $\calR(S\cup H')=a$. For a monotone rule, we can assume w.l.o.g. that all voters in $H'$ voter $a$ and thus the definitions coincide. 

Values $\live<1$ correspond to a \emph{stronger} liveness requirement, whereas $\live>1$ is relaxing liveness. 

 Note that values $\range>1$ in the definition of the outcome range (Def.~\ref{def:range}) effectively mean that we may replace all honest voters and, furthermore, add additional $(1-\range)|H|$ voters.

\begin{theorem}[Approximate Liveness]\label{thm:RE_maj_live}
  The $\tREMJ$ voting rule is $\live$-live if and only if 
  $$\live>\frac{(1-\mu)(1+\tau)}{2(1-\sigma-\mu)}.$$
\end{theorem}

\begin{proof}
Since any vote for $r$ reduces liveness, w.l.o.g\ all voters vote for $r$. 
There are $h^+= 1-\mu-\sigma$ active honest voters (all vote for $r$) . 
Suppose we  create a new profile $\ol V$ by moving a fraction of $\live$ votes from $r$ to $p$, then $p$ has $\ol v_p^+ = \ol h^+_p = \live(1-\mu-\sigma)$ votes.

In contrast, $r$ has $\ol h^+_r = h^+ - \ol h^+_p =  1-\mu-\sigma-\ol h^+_p$ active honest  votes remaining, plus $\sigma$ sybils. The $\tREMJ$ mechanism  adds $\tau(1-\mu)$ votes  so the total support for $r$ is 
$$\ol v_r^+ = (1-\mu-\sigma-\ol h^+_p)+\sigma+\tau(1-\mu)= (1+\tau)(1-\mu)-\ol h^+_p\ .$$

Since liveness requires $\ol v_p^+ > \ol v_r^+$, 
we get a tight bound of $2\ol h^+_p > (1+\tau)(1-\mu)$, or, equivalently, 
$$\live = \frac{\ol h^+_p}{1-\mu-\sigma} > \frac{(1+\tau)(1-\mu)}{2(1-\mu-\sigma)}\ ,$$
as required.
\end{proof}

\section{Mean Function}\label{section:mean}

One natural aggregation function in $\mathbb R^d$ is the mean function $\calG(V)=\frac{1}{|V|}\sum_{i\in V}s_i$.


If we assume the domain is unbounded then the questions of safety and liveness are moot, because every single voter (honest or sybil) can arbitrarily determine the location of the mean, regardless of the profile.

Let us assume then that the domain is $[0,1]^d$. Note that it matters where we set $r$.

Since there are already many parameters, we will consider the questions of sybils and partial participation separately. 
First, $\tREMN$ cannot guarantee $0$-safety even in the presence of a small fraction of sybils. 

\begin{proposition}
$\tREMN$ is not $\safe$-safe w.r.t. the mean for any $\safe<\frac{\sigma}{1+\tau}$. This is true regardless of $r$.
\end{proposition}
\begin{proof}
It is enough to consider a single dimension, where all honest voters are on $r$, and all sybils are on $1$. Since we will not use negative locations, we normalize the interval so that $r=0$. Then $\tREMN(V)=\frac{\sigma}{1+\tau}>0=\MN(H)$.
The highest we can push the outcome in $H'$ is by moving $\safe$ voters from $r=0$ to $1$, but
$$\MN(H')\leq (1-\safe)0+\safe 1 =\safe < \frac{\sigma}{1+\tau} = \tREMN(V),$$
So $\tREMN(V)\notin \calB(r;\ol{\MN}_\safe(H))$.
\end{proof}



Note that we cannot guarantee $0$-safety: if the honest voters are on $r$ and the sybils are not, then any number of virtual voters on $r$ will not cancel out the sybils. 

On the other hand, a mechanism that removes the $\tau$-most extreme voters (similarly to $\tSM$) is $0$-safe if $\tau\geq \sigma$.

\section{Nonatomic Population}\label{sec:rand_nonatomic}

We consider a nonatomic population of voters, which can be thought of as the limit case of a large population. In this case, we only care about the \emph{fraction} of voters for each alternative, and we can assume that under random participation,  this fraction is exactly the same among passive and active honest voters. We can see this in Figure~\ref{fig:binary_MJ}, where the distribution of honest voters (in blue) under random participation is much more balanced than under arbitrary participation.  This will allow us to show an improved safety-liveness tradeoff. 

\full{Fig.~\ref{fig:binary_MJ}(a) shows an example where there is a large majority of honest voters for $r$, and yet $\tSQMJ$ selects $p$. Thus, this profile implies a violation of $\safe$-safety whenever $\safe < \frac{h_r-h_p}{2}$.
Otherwise, we can define an profile $H'$ where $\safe|H|$ honest voters switch from $r$ to $p$ and get $\MJ(H'\cup S)=p$.
}

\full{
\begin{figure}
    \centering
    \includegraphics[width=0.8\textwidth]{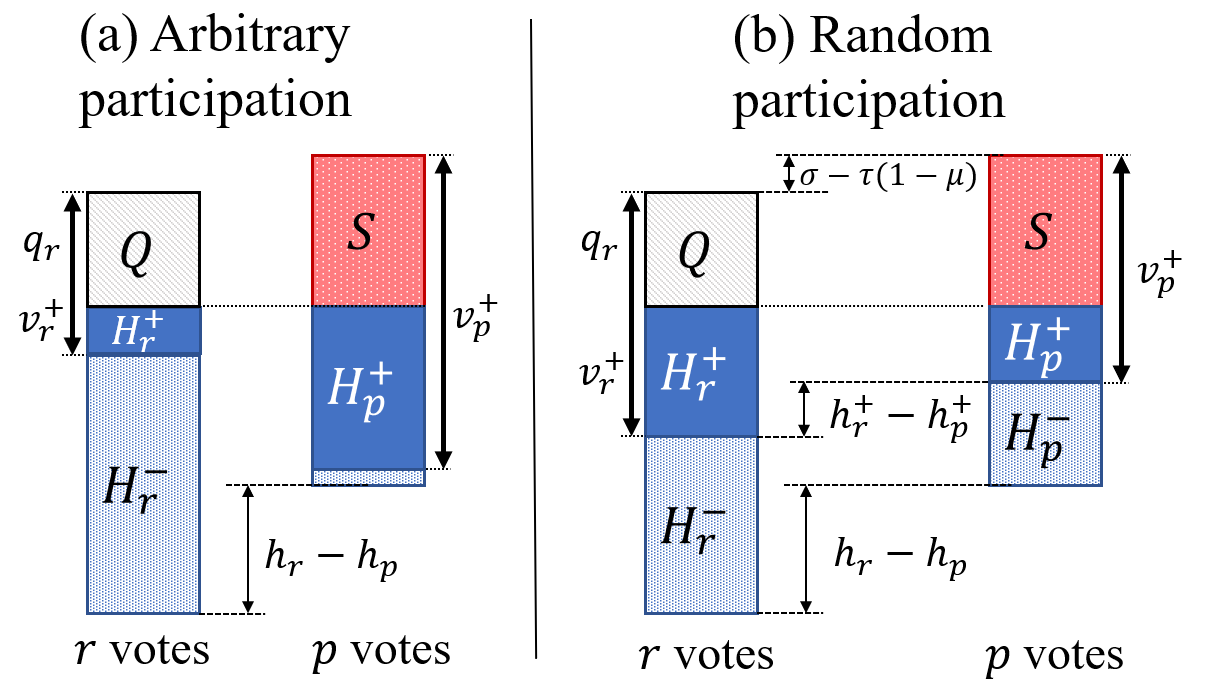}
    \caption{\label{fig:binary_MJ}An example of  voting profiles with the same $\sigma,\mu$ parameters under arbitrary partial participation (a), and under random partial participation (b). The thick arrows show the total amount of active votes for each alternative.}
\end{figure}
}

\begin{theorem}\label{thm:RE_maj_random_safe}
 For a nonatomic population with random participation, the $\tSQMJ$ voting rule is $\safe$-safe w.r.t Majority as the base rule if and only if $\safe \geq \frac{(\sigma - \tau(1 - \mu))(1-\sigma)}{2(1 - \mu - \sigma)}$.
\end{theorem}

\begin{proof}
Recall we denote by $u_r,u_p$ the fraction of voters for $r$ and $p$, respectively, in a voter set $U$. 

Suppose first that $h_p > h_r -2\safe$. This means that there is a profile $H'$ where we move only $\safe h$ voters from $r$ to $p$, and $\MJ(H')=p
$. Thus $p\in \ol{MJ}_\safe(H)\subseteq \calB(r;\ol{MJ}_\safe(H))$, which means $\safe$-safety holds.




The fraction of active voters among $H$ is denoted by  $\phi:=\frac{|H^+|}{|H|}=\frac{1-\mu-\sigma}{1-\sigma}$. 

Therefore:
  \labeq{eps_rand_inf}
  {h^+_r-h^+_p =  \phi(h_r-h_p)}

By definition,  $\tSQMJ(V^+) = \MJ(H^+ \cup S \cup Q)$ where $Q$ contains $\tau(1-\mu)$ voters for $r$.

Thus the total fraction of active $r$ voters is at least $h^+_r+\tau(1-\mu)$ (see Fig. 3(b)). As in the previous proofs, w.l.o.g. all sybils vote for $p$ as this is the worst case for safety. We get that
\begin{align*}
v^+_r&-v^+_p \geq (h^+_r + \tau(1-\mu)) - (h^+_p + \sigma)\tag{equality when $s_p=\sigma$}  \\
&=   (h^+_r-h^+_p) - (\sigma- \tau(1-\mu))\\
&=    \phi(h_r-h_p)  - (\sigma- \tau(1-\mu)) \tag{by Eq.~\eqref{eq:eps_rand_inf}}\\
&> \frac{1-\mu-\sigma}{1-\sigma}\left( \frac{(\sigma - \tau(1 - \mu))(1-\sigma)}{1 - \mu - \sigma} \right) - (\sigma- \tau(1-\mu))\\
&=  (\sigma- \tau(1-\mu))   - (\sigma- \tau(1-\mu))\\
&=0 ,
\end{align*}
as required. 

\medskip
In the other direction (i.e. to show tightness of the bound), consider any profile where all sybils vote $p$, and we set $h_p$ such that the equation 
holds with reversed inequality. That is,

\labeq{beta_rand_inf_rev}
{\frac{(\sigma - \tau(1 - \mu))(1-\sigma)}{2(1 - \mu - \sigma)} > \frac{h_r-h_p}{2}.}

Then the inequalities in the last block of equations are reversed and we get that $v^+_r-v^+_p<0$, meaning $\tSQMJ(V^+)=p$.

On the other hand, for any $\safe>\frac{h_r-h_p}{2}$, we have that $\ol{\MJ}_\safe(H)=\{r\}$ 

Joining both observations,   $\tSQMJ$ is not $\safe$-safe for any value of $\safe$ in the range $(\frac{h_r-h_p}{2},\frac{(\sigma - \tau(1 - \mu))(1-\sigma)}{2(1 - \mu - \sigma)})$.
\end{proof}

Random participation does not allow us to improve the bound on liveness beyond  Theorem~\ref{thm:RE_maj_live}, which is still tight. 

As a result of Theorem~\ref{thm:RE_maj_random_safe}, we get a better safety-liveness tradeoff under random participation:
\begin{corollary}\label{cor:REMJ_random}
  Under a nonatomic population with random participation:
  \begin{itemize}
      \item $\tSQMJ$ is safe w.r.t $\MJ$ iff $\tau \geq \frac{\sigma}{1 - \mu}$.
    \item $\tSQMJ$ is live iff $\tau< \frac{2(1-\sigma-\mu)}{1-\mu}-1$.
      \item We can get both if $3 \sigma + \mu < 1$.
  \end{itemize}
\end{corollary}

As with arbitrary participation, we show that the $\tSQMJ$ mechanism obtains the best possible tradeoff. 

\begin{theorem}[Lower bound for random participation]\label{thm:rand_LB}
Under random participation and nonatomic population, there is no rule $\calR$ such that $\calR^+$ is both $0$-safe and $1$-live when $3\sigma+\mu\geq 1$.
\end{theorem}

\begin{proof}
We denote by $\phi:=\frac{|H^+|}{|H|}=\frac{1-\mu-\sigma}{1-\sigma}$ the fraction of active honest voters. 

Suppose the mechanism is 1-live. 
By 1-liveness, there is a profile $V$ s.t. all sybils are voting for $r$, and $\calR^+(V)=\calR(S\cup H^+)=p$.

For a nonatomic population, $h^+_p=\phi h_p$ exactly.

 

Now, consider a profile $\ol V=\ol S\cup \ol H^+ \cup \ol H^-$, where $|\ol S|=|S|, |\ol H^+|=|H^+|, |\ol H^-|=|H^-|$, so $\sigma$ and $\mu$ are the same as in $V$. As in the proof of Thm~\ref{thm:arb_LB}, set $\ol s_p:=\min\{h^+_p,\sigma\}$ sybils to vote for $p$. The difference from Thm.~\ref{thm:arb_LB} is that we cannot set $\ol h^+_p$ directly (since they are selected at random), only $\ol h_p$. We set  
\labeq{ol_hp}
{\ol h_p:=h_p-\frac{\ol s_p}{\phi}.}
All other voters vote for $r$. 

Now, note that the total amount of active $p$ voters is 
$$\ol v^+_p = \ol s_p + \ol h^+_p = \ol s_p + \phi \ol h_p = \ol s_p + \phi (h_p-\frac{\ol s_p}{\phi}) = \ol s_p - \ol s_p +\phi h_p = \phi h_p = h^+_p = v^+_p.$$

This means that (as in Thm.~\ref{thm:arb_LB}), profiles $V$ and $\ol V$ are indistinguishable, and $\calR^+(\ol V)=\calR^+(V)=p$.  

We still need to show that $\ol h_r\geq \ol h_p$, which entails a violation of $0$-safety. 
Assume first that $\ol s_p < h^+_p$. Then $\ol s_p=\sigma$ and:
\begin{align*}
    \phi(\ol h_r -\ol h_p) &=  - \phi(\ol h-2\ol h_p) = \phi(1-\sigma-2\ol h_p) \\
    &= \phi(1-\sigma - 2(h_p-\frac{\ol s_p}{\phi})) = \phi(1-\sigma) -2\phi h_p + 2\ol s_p \tag{By Eq.~\ref{eq:ol_hp}}\\
    &= \frac{1-\sigma-\mu}{1-\sigma}(1-\sigma) -2 h^+_p +2\ol s_p\tag{By def. of $\phi$}\\
    &= 1-\sigma-\mu +2\sigma -2h^+_p \tag{as $\ol s_p= \sigma$}
    = 1+\sigma-\mu-2h^+_p \\
    &\geq 1+\sigma -\mu -2h^+ \geq 1+\sigma -\mu -2(1-\sigma-\mu)\\
    &= 3\sigma+\mu-1 \geq 0,
\end{align*}
where the last inequality is by the premise of the theorem. 
Since $\phi>0$, this entails $\ol h_r -\ol h_p\geq 0$ as well.

If $\ol s_p=h^+_p$ then 
$$\ol h_p = h_p-\frac{\ol s_p}{\phi} = h_p-\frac{h^+_p}{\phi} = h_p-h_p = 0,$$
meaning that in $\ol V$ there are no honest voters for $p$. In particular $\ol h_r > 0 = \ol h_p$. 
\end{proof}

\subsection{A General Result about Homogeneous Rules}

A voting rule $\calR$ is \emph{homogeneous} if $\calR(\alpha V)=\calR(V)$ for all $\alpha>0$. Note that majority, mean, median, etc. all homogeneous. 
\begin{proposition} \label{prop:hom_safe_UB}
With continuous population, every homogeneous rule $\calG$ is $\max\{\frac{\mu}{1-\sigma},\frac{\sigma}{1-\mu}\}$-safe with respect to itself.
\end{proposition}
\begin{proof}
Suppose first that $\frac{\sigma(1-\sigma)}{1-\mu}\geq \mu$, and let $\mu':=\frac{\sigma(1-\sigma)}{1-\mu}- \mu$.

We define $H'$ follows:  Selecting all of $H^-$, and additional $\mu'$ voters from $H^+$. These are $\frac{\sigma(1-\sigma)}{1-\mu}$ selected voters in total. 
Assign all of them uniformly to the locations of $S$.
Denote the new locations by $H'_S$ and the unchanged part of the profile by $H'_H$.

By construction,  $H'_S=xS$ and $H'_H=yH^+$ for some $x,y$. We need to verify that $x=y$. Indeed, 
$$x=\frac{|H'_S|}{|S|}=\frac{\frac{\sigma(1-\sigma)}{1-\mu}}{\sigma}= \frac{1-\sigma}{1-\mu},$$
whereas
$$y=\frac{|H'_H|}{|H^+|
}= \frac{|H|-|H'_S|}{1-\mu-\sigma}=\frac{1-\sigma-\frac{\sigma(1-\sigma)}{1-\mu}}{1-\mu-\sigma}= \frac{(1-\sigma)(1-\frac{\sigma}{1-\mu})}{1-\sigma-\mu}= \frac{(1-\sigma)\frac{1-\mu-\sigma}{1-\mu}}{1-\sigma-\mu}=\frac{1-\sigma}{1-\mu}.$$

Therefore $H'= \frac{1-\sigma}{1-\mu}V$, and due to homogeneity
 $$\calG(V)=\calG(H')\in \calB(r;\ol \calG_\sigma(H)).$$

The relative fraction of voters we moved is 
$$\safe = \frac{|H'_S|}{|H|}=\frac{\frac{\sigma(1-\sigma)}{1-\mu}}{1-\sigma}=\frac{\sigma}{1-\mu}.$$

If $\frac{\sigma(1-\sigma)}{1-\mu}< \mu$,
then we reassign the selected voters $H'_S\subseteq H^-$ in the same way over $S$. Then we reassign the remaining  $\mu-\frac{\sigma(1-\sigma)}{1-\mu}$ voters of $H^-$ over $H^+$.
One can check that $H'= \frac{1-\sigma}{1-\mu}V$ as in the previous case. The difference is that we moved $\mu = \frac{\mu}{1-\sigma}|H|$ voters so $\safe =  \frac{\mu}{1-\sigma}$.
\end{proof}


\end{document}